\documentclass[10pt,aps,prd,twocolumn,showpacs,superscriptaddress,nofootinbib,nobibnotes,longbibliography,floatfix]{revtex4-2}

\usepackage{bm}
\usepackage{mathtools}
\usepackage{amsmath}
\usepackage{amssymb}
\usepackage{amsfonts}
\usepackage{mathrsfs}
\usepackage{graphicx}
\usepackage{tensor}
\usepackage{accents}
\usepackage{commath}
\usepackage{chngcntr}
\usepackage{siunitx}
\usepackage[dvipsnames]{xcolor}
\usepackage[unicode]{hyperref}
\hypersetup{colorlinks=true, citecolor=MidnightBlue,
            linkcolor=MidnightBlue, urlcolor=MidnightBlue, linktocpage=true}
\usepackage[normalem]{ulem}
\usepackage{journals}
\usepackage{multirow}

\usepackage{orcidlink}
\newcommand{\orcid}[1]{\href{https://orcid.org/#1}{\includegraphics[width=10pt]{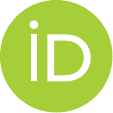}}}

\definecolor{darkgreen}{rgb}{0.0,0.6,0}

\newcommand{\etal}{\emph{et al.}}

\begin{document}

\title{Impact of the relativistic Cowling approximation on shear and interface modes of neutron stars}

\author{Christian J. Kr\"uger \orcid{0000-0003-2672-2055}}
    \email{christian.krueger@tat.uni-tuebingen.de}
    \affiliation{Theoretical Astrophysics, IAAT, University of T\"ubingen, 72076 T\"ubingen, Germany}
\author{Hajime Sotani \orcid{0000-0002-3239-2921}}
    \email{sotani@yukawa.kyoto-u.ac.jp}
    \affiliation{Department of Mathematics and Physics, Kochi University, Kochi, 780-8520, Japan}
    \affiliation{Astrophysical Big Bang Laboratory, RIKEN, Saitama 351-0198, Japan}
    \affiliation{Interdisciplinary Theoretical and Mathematical Science Program (iTHEMS), RIKEN,
Saitama 351-0198, Japan}
    \affiliation{Theoretical Astrophysics, IAAT, University of T\"ubingen, 72076 T\"ubingen, Germany}

\date{\today}

\begin{abstract}
We investigate shear and interface modes excited in neutron stars with an elastic crust in the full general relativistic framework and compare them to the results obtained within the relativistic Cowling approximation. We observe that the Cowling approximation has virtually no impact on the frequencies or the eigenfunctions of the shear modes; in contrast, the interface modes that arise due to the discontinuities of the shear modulus experience a considerable shift in frequency when applying the Cowling approximation . Furthermore, we extend a scheme based on the properties of the phase of amplitude ratios, which allows us to estimate the damping times of slowly damped modes; our extension can provide an estimation of the damping time even if the features of the amplitude ratio are incomplete or if some of them violate the underlying linearity assumption. The proposed scheme is also computationally less expensive and numerically more robust and we provide accurate estimates as well as lower bounds for damping times of shear and interface modes. We estimate the damping times also via the quadrupole formula and find that it provides good order-of-magnitude estimates.
\end{abstract}

\maketitle

\section{Introduction}

Neutron stars (NSs) provide an environment to probe matter under extreme physical conditions that are unrivaled by terrestrial laboratories; it is assumed that the densities in the core of NSs exceed nuclear saturation density \cite{Shapiro:1983du}. Due to their high compactness and resulting strong gravitational fields, Einstein's theory of gravity is essential for their accurate description. A crucial ingredient for the description of the interior of a neutron star is the high-density equation of state (henceforth EOS), which is, hitherto, afflicted with large uncertainties. Constraining the nuclear EOS is one of the most pressing efforts of the astrophysical community at the present time.

Crucial in ruling out a large number of proposed EOSs are the discoveries of $2 M_\odot$ neutron stars~\cite{2010Natur.467.1081D, 2013Sci...340..448A, 2020NatAs...4...72C, 2021ApJ...915L..12F} that impose a lower limit on the neutron star maximum mass. Recent observations in the electromagnetic spectrum by the NICER telescope have provided constraints on mass and radius of the pulsars PSR J0030+0451 \cite{2019ApJ...887L..21R, 2019ApJ...887L..24M} and PSR J0740+6620 \cite{2021ApJ...918L..27R, 2021ApJ...918L..28M} by modeling their hotspots. Incorporating observations in the gravitational wave sector also opens up entirely new avenues toward new neutron star measurements and is part of the new, exciting field of multi-messenger astronomy. The first observation of gravitational waves from the binary neutron star merger GW170817 \cite{LIGOScientific:2017vwq, LIGOScientific:2017ync} enables us to put constraints on the tidal deformability and radius of neutron stars with a mass of $1.4 M_\odot$ \cite{2017ApJ...850L..34B, 2018PhRvL.120q2703A, 2018PhRvL.120z1103M, LIGOScientific:2018cki}. Other observations of importance were the detections of quasi-periodic oscillations (QPOs) in the X-ray tails of giant magnetar flares, such as in SGR 0526-66 \cite{1983A&A...126..400B}, SGR 1900+14 \cite{2005ApJ...632L.111S}, SGR 1806-20 \cite{2005ApJ...628L..53I, 2006ApJ...637L.117W, 2006ApJ...653..593S}, or GRB 200415A \cite{2021Natur.600..621C}; such oscillations have also been detected in the pre-merger stage preceding GRB 211211 \cite{2024ApJ...970....6X}. Although the precise nature of these QPOs, which have frequencies around a few tens of Hz, is still an open question, they are often linked to crustal oscillations of the star.

Stellar oscillations depend sensitively not only on the neutron star's bulk quantities but also on its interior composition, and their observation via their imprint on gravitational waves opens up the field of \emph{gravitational wave asteroseismology} \cite{1998MNRAS.299.1059A}, but, as stated above, such oscillations may also impact electromagnetic signals. In asteroseismology, measured oscillation frequencies and/or damping times are translated into information on the neutron star's characteristics \cite{1996PhRvL..77.4134A, 2004PhRvD..69h4008S, 2012MNRAS.419..638P, 2013PhRvD..88d4052D}; due to the observed QPOs which are likely related to crustal oscillations, there are several studies which are concerned with extracting information from such oscillations \cite{2007MNRAS.375..261S, 2011MNRAS.414.3014C, 2012PhRvL.108t1101S, 2024Univ...10..231S}.

A precise understanding of the spectrum of neutron stars is, hence, crucial to extracting information from current as well as potential future observations. Their high compactness calls for a general relativistic treatment which considerably complicates the perturbation equations that need to be solved and there are numerous publications approaching this problem, employing different approximations or techniques \cite{1967ApJ...149..591T, 1983ApJS...53...73L, 1985ApJ...292...12D, 1992MNRAS.255..119K, 1995MNRAS.274.1039A, 1998PhRvD..58l4012A, 2002MNRAS.332..676R, 2002MNRAS.334..933J, 2007PhRvD..76j4033F, 2020PhRvD.102f4026K}. One of the most commonly applied approximations is the so-called \emph{Cowling approximation}, in which perturbations of the spacetime are neglected, resulting in a considerable simplification of the equations \cite{1941MNRAS.101..367C, 2001MNRAS.325.1463F, 2003MNRAS.339.1170R, 2007PhRvD..75d3007B, 2008PhRvD..77l4019K, 2008PhRvD..78f4063G, 2024arXiv240920178C}. The Cowling approximation is well-known for overestimating the frequency of the fundamental ($f$-) mode and the pressure-restored (acoustic) ($p$-) modes \cite{1997MNRAS.289..117Y, 2001MNRAS.322..389Y}.

The spectrum of neutron stars is very rich and a multitude of different oscillation modes may be excited due to various astrophysical processes. In this study, we focus particularly on the shear ($s$-) and interface ($i$-) modes, which are associated with the presence of an elastic crust \cite{1990MNRAS.245...82F, 2002A&A...395..201Y, 2015PhRvD..92f3009K, 2017PhRvC..96f5803F}. It is expected that the Cowling approximation has no significant impact on these oscillation modes, however, there is, to the best of our knowledge, no quantitative study on this topic. We aim to fill this gap with the present investigation. We assume that the neutron star is old and cold and that the crust has crystallized across its entire extent \cite{1993PhRvE..47.4330F}; the core and the envelope of the neutron star are modeled as a perfect fluid. We use our respective numerical codes in full General Relativity as well as in the relativistic Cowling approximation (details on their implementation can be found in Refs. \cite{krueger_thesis, 2023PhRvD.107l3025S} and references therein) to investigate the stellar spectrum of a few selected neutron stars models with a focus on crustal oscillations. We then scrutinize the impact of the Cowling approximation on the $s$- and $i$-modes; at the same time, we also determine or estimate the damping times of these modes. Our study provides insight into which oscillation modes of neutron stars may safely be investigated within the much simpler and computationally less expensive Cowling approximation or in which cases a full General Relativistic treatment may be advantageous.

The layout of this paper is as follows. In Sec.~\ref{sec:eos_equil}, we show the equations we solve in order to construct equilibrium models and introduce the employed equation of state along with the shear modulus of the elastic crust. In Sec.~\ref{sec:pert_eq}, we explain the underlying definitions for the perturbation equations, both in the Cowling approximation as well as in full General Relativity. We devote Sec.~\ref{sec:damp_times} to two methods that allow determining the damping times of long-lived modes. In the following Sec.~\ref{sec:results}, we discuss the eigenfrequencies and eigenfunctions of the shear and interface modes and compare their results obtained in the Cowling approximation and in full general Relativity; we also determine values or constraints for their damping times. We conclude the study in Sec.~\ref{sec:conclusions}. Unless otherwise mentioned, we use units in which $c = G = 1$, where $c$ and $G$ denote the speed of light and the gravitational constant, respectively.

\section{EOS and Equilibrium Models}
\label{sec:eos_equil}

\subsection{General Setup}

As this study is a continuation of prior studies, cf. Refs.~\cite{2023PhRvD.107l3025S, 2024PhRvD.109b3030S}, we consider non-rotating and unstrained NS models as equilibrium configurations here. Such objects can be described by the metric, $g_{ab}$, given by the line element
\begin{equation}
    \dif s^2
    = - e^{2\Phi} \dif t^2 + e^{2\Lambda} \dif r^2 + r^2 \left( \dif\theta^2 + \sin^2\theta \dif\phi^2\right),
    \label{eq:ds2_cowling}
\end{equation}
where $\Phi$ and $\Lambda$ are the metric functions depending only on $r$. A mass function $m(r)$ denoting the gravitational mass inside a ball of radius $r$ is defined as usual via
\begin{equation}
    e^{-2\Lambda} = 1 - \frac{2m(r)}{r}.
\end{equation}
Since we assume that the equilibrium configuration is unstrained, we employ the usual energy-momentum tensor of a perfect fluid for the NS matter, i.e.,
\begin{equation}
    T^{ab}_\textrm{pf}
    = (\epsilon + p) u^a u^b + p g^{ab},
\end{equation}
where $\epsilon$ is the energy density, $p$ the pressure, and $u^a$ the 4-velocity of the fluid element. We place the label ``$\textrm{pf}$'' (as a shortcut for perfect fluid) liberally downstairs or upstairs in order not to clutter the formulae.

The Einstein equations $G_{ab} = 8\pi T^\textrm{pf}_{ab}$ then imply the TOV equations
\begin{align}
    \Lambda'
    & = \frac{1 - e^{2\Lambda}}{2r} + 4\pi r e^{2\Lambda} \epsilon,
    \\
    \Phi'
    & = \frac{e^{2\Lambda} - 1}{2r} + 4\pi r e^{2\Lambda} p,
    \\
    p'
    & = - \left(\epsilon + p\right) \Phi',
\end{align}
where the prime denotes a derivative with respect to $r$. Provided an equation of state and a central pressure (or a central density), stellar models can be constructed by integrating the Tolman-Oppenheimer-Volkoff (TOV) equations, using standard Runge-Kutta solvers. 

To determine the stellar surface, one has to integrate the TOV equation up to the position, where the pressure becomes zero. But, since the pressure sharply drops down in the vicinity of the stellar surface, it is quite difficult to integrate the perturbation equations in such a region for solving the eigenvalue problem. So, we simply adopt $\rho_s=10^6$ g/cm$^3$ as a surface density (instead of a limiting pressure). In practice, the stellar mass, radius, and even the oscillation frequencies considered in this study are almost insensitive to the (non-zero) surface density, if it is small enough. 

On the other hand, to determine the frequencies of elastic oscillations focused on in this study, one has to know the region, where the elasticity is non-zero. The transition density between the crust and envelope (or the density at the crustal surface) significantly depends on the temperature, where the Fermi temperature becomes comparable to and less than the physical temperature~\cite{1983ApJ...272..286G}. So, in this study, we simply adopt $10^{10}$ g/cm$^3$ as the crust surface density as in Refs.~\cite{2023PhRvD.107l3025S, 2024PhRvD.109b3030S}. Meanwhile, the transition density for the inner boundary of the elastic region depends on the EOS, which will be mentioned below.

\subsection{Equation of State}
\label{sec:eos}

We employ a barotropic, zero temperature parametrized EOS based on nuclear parameters that have been proposed by Oyamatsu and Iida \cite{2003PThPh.109..631O, 2007PhRvC..75a5801O} (hereafter referred to as OI-EOS). The barotropicity of the EOS does not cause harm to the study as the oscillation modes under consideration, i.e. shear and interface modes, are virtually unaffected by this assumption. The bulk energy per nucleon is generally given as a function of the baryon number density, $n_\textrm{b}$, and an asymmetry parameter, $\tilde{\alpha}$,
\begin{equation}
    \frac{E}{A}
    = w_s(n_\textrm{b}) + \tilde{\alpha}^2 S(n_\textrm{b}) + \mathcal{O}\left(\tilde{\alpha}^3\right),
\end{equation}
where $n_\textrm{b} = n_\textrm{p} + n_\textrm{n}$ and $\tilde{\alpha} = \left(n_\textrm{n} - n_\textrm{p}\right) / n_\textrm{b}$ with the proton number density, $n_\textrm{p}$, and the neutron number density, $n_\textrm{n}$. The function $w_s$ describes the energy per nucleon of symmetric nuclear matter ($\tilde{\alpha} = 0$), and $S$ denotes the density-dependent symmetry energy. These two functions can be expanded around the saturation density of symmetric nuclear matter, $n_0 \approx 0.15 - 0.16\,\textrm{fm}^{-3}$ \cite{2017RvMP...89a5007O}, as a function of $u = (n_\textrm{b} - n_0)/(3n_0)$:
\begin{align}
    w_s(n_\textrm{b})
    & = w_0 + \frac{K_0}{2} u^2 + \mathcal{O}\left( u^3 \right),
    \\
    S(n_\textrm{b})
    & = S_0 + L u + \mathcal{O}\left( u^2 \right).
\end{align}
A particular EOS is then fixed after providing the five coefficients $n_0$, $w_0$, $K_0$, $S_0$, and $L$. Of those five parameters, the three parameters $n_0$, $w_0$, and $S_0$ are relatively well constrained via terrestrial experiments; these are the nuclear saturation density $n_0$ which is determined by $\partial w_s/ \partial u = 0$, and the binding energy $w_0$ of symmetric matter and the symmetry energy $S_0$ which are both determined at saturation density $n_\textrm{b} = n_0$. The higher order expansion parameters $K_0$ and $L$ are more difficult to constrain in laboratory experiments; the current fiducial ranges for these parameters are $K_0 = 240 \pm 20\,\textrm{MeV}$~\cite{2006EPJA...30...23S} and $L = 60 \pm 20\,\textrm{MeV}$~\cite{2014EPJA...50...27V, 2019EPJA...55..117L}. The values for the OI-EOS adopted in this study are $n_0 = 0.1598\,\textrm{fm}^{-3}$, $w_0 = -16.15\,\textrm{MeV}$, $S_0 = 31.08\,\textrm{MeV}$, $K_0 = 230\,\textrm{MeV}$, and $L = 42.59\,\textrm{MeV}$, which are inside the fiducial ranges of $K_0$ and $L$.

Further, in order to investigate the dependence of our results on the EOS stiffness in the high-density region, we also consider a modification to the OI-EOS at densities larger than $\epsilon_t = 0.536 \times 10^{15} \textrm{g}/\textrm{cm}^3$; this energy density corresponds to a number density of $2n_0$, i.e. twice the nuclear saturation density. For densities larger than $\epsilon_t$, we replace the OI-EOS with a one-parameter EOS given by
\begin{equation}
    p = \alpha \left(\epsilon - \epsilon_t \right) + p_t,
\end{equation}
where $p_t = p(\epsilon_t)$ is the pressure corresponding to the energy density $\epsilon_t$ according to the OI-EOS. The speed of sound for this high-density EOS is simply $c_s^2 = \alpha$. In particular, we use the values $\alpha = 1/3, 0.6,$ and $1$ in this study as in Refs.~\cite{2023PhRvD.107l3025S, 2024PhRvD.109b3030S}.

\subsection{Shear Modulus}
\label{sec:shearmodulus}

As we are interested in shear modes, which are the oscillations inherent to an elastic crust, we need to specify the shear modulus, $\mu$. This quantity will assume non-zero values only in the crystallized crust region, which roughly spans the densities from $10^{10}\,\textrm{g}/\textrm{cm}^3$ to $10^{14}\,\textrm{g}/\textrm{cm}^3$. Once a non-zero shear modulus is employed, the elastic crust will enrich the stellar spectrum by $s$-modes (shear modes) in the polar sector and by $t$-modes (torsional modes) in the axial sector; additionally, the discontinuous change of the shear modulus, e.g., at the envelope-crust interface and the crust-core interface, gives rise to polar $i$-modes (interface modes).

Most parts of the crust region are composed of spherical nuclei, while non-spherical nuclei would also appear at the basis of the crust (above the core region.)  Since the shear modulus in the phase of slablike nuclei, $\mu_\textrm{sl}$, can be neglected at linear perturbation level, i.e., $\mu_\textrm{sl} = 0$ \cite{1998PhLB..427....7P}, we consider the elastic oscillations excited only in the regions composed of spherical or cylindrical nuclei in this study.
We distinguish between two shear moduli depending on whether the nuclei in the crust are of spherical or cylindrical shape. For the region of the crust that is composed of spherical nuclei, we adopt a shear modulus that assumes that the nuclei form a body-centered cubic (bcc) lattice with a pointlike ion. The shear modulus of zero-temperature matter, $\mu_\textrm{sp}$, is then a function of the ion charge number $Z$, the ion number density $n_i$, and the Wigner-Seitz cell radius $a$ (related to $n_i$ via $4\pi a^3/3 = 1/n_i$) and is given by \cite{1991ApJ...375..679S}
\begin{equation}
    \mu_\textrm{sp}
    = 0.1194 \frac{n_i \left( Ze \right)^2}{a}.
\end{equation}

Meanwhile, in the region where the nuclei take on a cylindrical shape, the shear modulus, $\mu_\textrm{cy}$, is given by \cite{1998PhLB..427....7P}
\begin{equation}
    \mu_\textrm{cy}
    = \frac{2}{3} E_\textrm{Coul}
        \times 10^{2.1(w_2 - 0.3)},
\end{equation}
where $E_\textrm{Coul}$ is the Coulomb energy per volume of a Wigner-Seitz cell and $w_2$ is the volume fraction of cylindrical nuclei.

The transition baryon number densities between the spherical and cylindrical nuclei, $n_{\rm SP/C}$ and between the cylindrical and slablike nuclei, $n_{\rm C/S}$, strongly depend on the adopted EOS, especially on the value of $L$ \cite{2003PThPh.109..631O, 2007PhRvC..75a5801O}. These values for the EOS considered in this study are respectively $n_{\rm SP/C}=0.06238$~fm$^{-3}$ and $n_{\rm C/S}=0.07671$~fm$^{-3}$.

\section{Perturbation Equations}
\label{sec:pert_eq}

In this study, we are comparing the frequencies of $s$- and $i$-modes of relativistic NS models in the full relativistic framework to those with the relativistic Cowling approximation. As usual, we linearise the Einstein equations in order to arrive at the differential equations that govern small perturbations around equilibrium. The relativistic Cowling approximation assumes that the spacetime is static and only the fluid undergoes oscillations; while in principle this can be achieved by nullifying the metric perturbations, i.e., $\delta g_{ab} \equiv\  h_{ab} = 0$, the derivation of the equations follows slightly different paths depending on whether the Cowling approximations is assumed or not. The equations that we solve for this study are well-known; we repeat the fundamental definitions here for the sake of clarity and completeness. See the last paragraph of this section for a summary of the publications in which the relevant equations are explained in more detail.

We define the Lagrangian displacement of the fluid in the fully relativistic case to be
\begin{align}
    \xi^i
    & = \left(
        e^{-\Lambda} \frac{\tilde{W}}{r},
        -\frac{\tilde{V}}{r^2} \frac{\partial}{\partial\theta},
        -\frac{\tilde{V}}{r^2} \frac{1}{\sin^2\theta} \frac{\partial}{\partial\phi}
        \right) r^l Y_{lm} e^{i\omega t}.
\intertext{while in the Cowling approximation the factor $e^{-\Lambda}$ and the $r^l$ dependence are absorbed in the functions $V$ and $W$:}
    \xi^i
    & = \left(
        rW, 
        V \frac{\partial}{\partial\theta},
        V \frac{1}{\sin^2\theta} \frac{\partial}{\partial\phi}
        \right) Y_{lm} e^{i\sigma t},
    \label{eq:def_xi_cowling}
\end{align}
where $V$, $\tilde{V}$, $W$, and $\tilde{W}$ are functions of $r$, $Y_{lm}(\theta, \phi)$ are the spherical harmonics (and we have omitted their angular dependence) and $\omega = \sigma + i/\tau$ is the complex-valued frequency. Since there is no damping due to gravitational wave emission in the Cowling approximation, the Lagrangian displacement in Eq.~\eqref{eq:def_xi_cowling} is separated into harmonics with respect to $\sigma = \textrm{Re}(\omega)$. As we are dealing with non-rotating stars, we may set $m = 0$.

In the fully relativistic case, we choose the well-known Regge-Wheeler gauge \cite{1957PhRv..108.1063R}, in which the metric perturbations $h_{ab}$ are defined as
\begin{equation}
h_{ab} = \begin{pmatrix}
e^{2\Phi} H_0 & r \dot{H}_1& 0 & 0 \\
r \dot{H}_1 & e^{2\Lambda} H_2 & 0 & 0 \\
0 & 0 & r^2 K  & 0 \\
0 & 0 & 0 & r^2 \sin^2\theta K
\end{pmatrix} r^l Y_{lm} e^{i\omega t},
\end{equation}
where $H_0$, $H_1$, $H_2$, and $K$ are functions of $r$ and the dot denotes the time derivative. The perturbation equations follow from the perturbed Einstein equations and conservation law for energy-momentum,
\begin{equation}
    \delta G_{ab} = 8\pi \delta T_{ab}
    \quad\text{and}\quad
    \delta \left( \nabla_a T^{ab} \right) = 0.
\end{equation}
In the Cowling approximation, all metric perturbations vanish and the perturbations are derived purely from the perturbed conservation law,
\begin{equation}
    \nabla_a \delta T^{ab} = 0,
\end{equation}
where we can pull the variation $\delta$ into the covariant derivative since the perturbations of the Christoffel symbols are zero.

Since we are concerned with oscillations of an elastic crust, we need to include shear stresses, described by the shear tensor $\Sigma_{ab}$, into the energy-momentum tensor $T_{ab}$ in addition to the perfect fluid part $T_{ab}^\textrm{pf}$. However, as we assume the background configuration to be unstrained, these shear stresses appear only in the perturbed configuration via $\delta \Sigma_{ab}$.

The shear tensor $\Sigma_{ab}$ is determined by
\begin{equation}
    \mathcal{L}_u \Sigma_{ab}
    = \frac{2}{3} \Sigma_{ab} \nabla_c u^c + \sigma_{ab},
\end{equation}
where $\mathcal{L}_u$ is the Lie derivative along the 4-velocity $u^a$ of the fluid and $\sigma_{ab}$ is the shear rate tensor given by
\begin{equation}
    \sigma_{ab}
    = \frac{1}{2} \left(
        \perp_a^{\ c} \nabla_c u_b + \perp_b ^{\ c} \nabla_c u_a
        \right)
        - \frac{1}{3} \perp_{ab} \nabla_c u^c,
\end{equation}
and $\perp_{ab}$ is the usual projection tensor:
\begin{equation}
    \perp_{ab} = g_{ab} + u_a u_b.
\end{equation}
Assuming a Hookean relationship \cite{1983MNRAS.203..457S} for the shear strain, the total perturbed energy-momentum tensor is
\begin{equation}
    \delta T_{ab}
    = \delta T_{ab}^\textrm{pf} - 2\mu \delta \Sigma_{ab},
\end{equation}
where $\mu$ is the shear modulus as defined in Sec.~\ref{sec:shearmodulus}. The perturbed shear strain tensor can be calculated to be
\begin{equation}
    \delta \Sigma_{ab}
    = \frac{1}{2} \left(
        \perp_a^{\ c} \perp_b^{\ d}
        - \frac{1}{3} \perp_{ab} \perp^{cd}
        \right) \Delta g_{cd},
\end{equation}
where $\Delta g_{ab} = h_{ab} + 2\nabla_{(a} \xi_{b)}$ are the Lagrangian metric perturbations.

With those ingredients, the perturbation equations concerning the interior of the star can be derived. First, for the full general relativistic case, we show the perturbation equations valid in the elastic crust in Appendix~\ref{sec:perteq_elastic_fullgr}; the equations (and boundary conditions) relevant for a perfect fluid were first published in Ref.~\cite{1985ApJ...292...12D} but can also be found in Ref.~\cite{krueger_thesis}, where the definitions of the perturbation variables is identical to those in the present paper. At the boundaries of the elastic crust, where the shear modulus $\mu$ is discontinuous, we have to impose the continuity of the functions $\tilde{W}$ and $T_2:= \delta \Sigma_r^{\ \theta}$ as well as the metric functions $H_0$, $H_1$, and $K$ as junction conditions~\cite{2011PhRvD..84j3006P, krueger_thesis} owing to the continuity of the first and second fundamental forms; additionally, at the interface between spherical and cylindrical nuclei within the crust, where the shear modulus $\mu$ exhibits a small jump, too, we impose continuity of the Lagrangian pressure perturbation, $X$, given in Appendix \ref{sec:perteq_elastic_fullgr}. In the exterior of the star, we solve the Zerilli equation using the method employing complex coordinates as detailed in Ref.~\cite{1995MNRAS.274.1039A}. Second, for the perturbation equations as well as boundary and junction conditions and the corresponding in the Cowling approximation, we refer the reader to Apps. A to C in Ref.~\cite{2023PhRvD.107l3025S}, where the solution strategy is explained in detail.

\section{Damping Times of Slowly Damped Modes}
\label{sec:damp_times}

We perform our calculations of the oscillation modes in the frequency domain; in the full general relativistic case, this approach yields, in principle, the frequency as well as the damping time at the same time. This method works well for $w$-modes \cite{1992MNRAS.255..119K} and also for the $f$-mode and low-order $p$-modes \cite{1983ApJS...53...73L, 1995MNRAS.274.1039A}. However, due to finite numerical accuracy and truncation errors, determining the damping time of very slowly damped modes (for which the imaginary part of the complex-valued frequency is several orders of magnitude smaller than their real part) is computationally challenging and a direct iteration in the complex plane for a zero of $A_\textrm{in}(\omega)$ sometimes impossible. The frequency of such modes can be estimated accurately by analysing the amplitude of the ingoing wave $A_\textrm{in}(\omega)$ for purely real-valued frequencies $\omega$; in the following, we discuss to methods how estimates for the damping time can be found in these cases.

\subsection{Features of the Phase of the Amplitude Ratio}
\label{sec:damp_bw}

First, we discuss a method that has been devised by Andersson \emph{et al.}~\cite{2002PhRvD..66j4002A}, which utilizes certain features of the ratio $\kappa := A_\textrm{out} / A_\textrm{in}$ of the outgoing and ingoing wave amplitudes for purely real-valued frequencies $\omega$. We extend this method in order to also be able to yield constraints on the damping time in the case that not all desired features can be resolved. As this method is rather technical, we show it including our extensions in full detail in App.~\ref{app:constrain_damping}, and repeat only the main results here. In our tests, this method yields very accurate values for the damping times of $f$- and $p$-modes (which, in many cases, can also be determined by direct iteration in the complex plane), proving its reliability. As an example, we detail one such test in App.~\ref{app:kappa_test}.

The method investigates the quantity $\textrm{Im}(\kappa) / \textrm{Re}(\kappa)$, which is closely linked to the phase of $\kappa$; it can be argued that this quantity features two poles (denoted by $P_1$ and $P_2$) and two zeroes (denoted by $Z_1$ and $Z_2$) in the vicinity of an actual eigenfrequency $\omega_n$ of the system (cf. Fig.~\ref{fig:s1_breit_wigner}, in which two poles and one zero associated with an $s$-mode can be seen). In the optimal case, both poles and both zeroes can be located and the damping time is then given by
\begin{align}
    \tau^{-1}
    & = \sqrt{(\alpha_n - P_1)(P_2 - \alpha_n)} \\
    & = \sqrt{(\alpha_n - Z_1)(Z_2 - \alpha_n)},
\end{align}
where $\alpha_n$ is the frequency of the mode. This can be determined as the minimum of the Breit-Wigner resonance or by the estimate
\begin{align}
    \alpha_n
    & = \frac{P_1 P_2 - Z_1 Z_2}{P_1 + P_2 - Z_1 - Z_2}.
    \label{eq:alpha_n}
\end{align}
It is immediately clear that (mathematically) the knowledge of either the two poles or the two zeroes suffices to determine the damping time $\tau$ when $\alpha_n$ can be determined accurately without relying on Eq.~\eqref{eq:alpha_n}.

In our analyses, we observed that sometimes the location of $\alpha_n$ relative to the poles or zeroes could not be accurately enough determined to avoid using Eq.~\eqref{eq:alpha_n}; this may happen since locating a minimum can numerically not be as accurate as locating a pole or zero. Hence, it is useful to also be able to provide constraints on the damping time. We find that
\begin{align}
    |Z - P|
    < \tau^{-1} \lesssim 1.2 |Z - P|;
\end{align}
in this formula, $P$ and $Z$ denote the one pole and the one zero directly adjacent to the mode frequency $\alpha_n$. This means that by simply locating the closest pole and zero, we can estimate the damping time to an error of less than 20\%. Second, we can derive the estimate
\begin{align}
    0.70 \delta \lesssim 2 \tau^{-1} < \delta,
\end{align}
where $\delta$ is the smaller of the differences between the poles and the zeroes, i.e., $\delta := \min(P_2 - P_1, Z_2 - Z_1)$. Last, in the case that a mode has such a large damping time that the poles and zeroes cannot be separated from each other at some frequency resolution $\Delta\omega$, then it is
\begin{align}
    \tau
    > 2 / \Delta\omega .
\end{align}

These are, in a compact way, the main results that we are able to extract from the features of the quantity $\textrm{Im}(\kappa) / \textrm{Re}(\kappa)$ and that allow us to either calculate or constrain the damping time $\tau$ of a mode. Nonetheless, we encourage the reader to delve into App.~\ref{app:constrain_damping} for a detailed derivation of these expression.

\subsection{Quadrupole formula}
\label{sec:damp_qf}

Other than using the features of phase of the amplitude ratio, we can also estimate the damping time of slowly damped modes by employing the well-known quadrupole formula \cite{1918SPAW.......154E}; it can then be calculated via
\begin{align}
    \tau
    & = - \frac{2 E}{\left< \dif E/\dif t \right>},
    \label{eq:damp_time_def}
\end{align}
where $E$ is the total energy stored in the oscillation mode and $\left< \dif E /\dif t \right>$ is the rate of energy loss due to gravitational wave emission. For the latter, no fully relativistic expression is known; hence, we follow the approach by Lioutas \& Stergioulas \cite{2018GReGr..50...12L} and employ their suggested ad-hoc modification of the Newtonian quadrupole formula as well as their denotation for the different quantities.

The energy $E$ of the mode is given by the maximum amplitude of the kinetic energy \cite{1969ApJ...158....1T, 2018GReGr..50...12L}
\begin{align}
    E
    & = \frac{1}{2} \omega_l^2 \int_0^R \left( \epsilon + p \right) r^{2l} \left(
        |\tilde{W}_l|^2 + l(l+1) |\tilde{V}_l|^2 
    \right) e^{\Lambda-\Phi} \dif r,
    \label{eq:E_kin_R}
\end{align}
where $\tilde{V}_l$ and $\tilde{W}_l$ are the eigenfunctions of the tangential and radial displacement of the corresponding oscillation mode which has the angular frequency $\omega_l$; this is a full general relativistic expression and is denoted with ``R''; its Newtonian counterpart, denoted with ``N'', can be obtained by replacing $(\epsilon + p) \rightarrow \rho$ and $e^{\Lambda-\Phi} \rightarrow 1$ in expression \eqref{eq:E_kin_R}.

The standard quadrupole formula (denoted ``SQF'') is a Newtonian expression and is given by \cite{1969ApJ...158..997T}
\begin{align}
- \left< \frac{\dif E}{\dif t} \right>
    & = \frac{2\pi (l+1)(l+2) }{l (l-1) [ (2l+1)!!]^2}
        \omega_l^{2l+2}
        \left( \int_0^R r^{l+2} \delta \rho_l \dif r \right)^2,
    \label{eq:sqf}
\end{align}
where $\delta \rho_l$ denotes the eigenfunction of the perturbed rest-mass density; as a full general relativistic formula is not known, we follow the suggestion of Ref. \cite{2018GReGr..50...12L} to substitute $\delta \rho_l \rightarrow \delta \epsilon_l$ (the Eulerian perturbation of the energy density) in Eq.~\eqref{eq:sqf}, which is then referred to as ``RQF''.

Altogether, we are provided with two expressions for both the energy $E$ and the rate of energy loss $\left< \dif E / \dif t \right>$, which we can plug into Eq.~\eqref{eq:damp_time_def}. We employ the two combinations N/SQF and R/RQF in order to calculate the damping time $\tau$ of a mode.

\section{Results}
\label{sec:results}

Having laid out the perturbation equations that need to be solved both in full General Relativity as well as in the Cowling approximation, we now turn to the analysis of the eigenfrequencies within those two formalisms. It is well-known that the frequencies of the fundamental ($f$-) and pressure-restored ($p$-)modes are considerably overestimated within the Cowling approximation \cite{1997MNRAS.289..117Y, 2001MNRAS.322..389Y} (see also Fig.~\ref{fig:s_modes}); hence, we will focus on the impact of the Cowling approximation on the $s$-modes that arise due to the elasticity of the crust and the two $i$-modes present in the stellar model considered in this study due to the two boundaries (where the shear modulus, $\mu$,  discontinuously becomes zero) of the elastic crust, i.e., the boundaries between the envelope 
and crust surface and between the phase of cylindrical nuclei and the phase of slablike nuclei (which is the crust-core interface).

In particular, we will investigate a number of NS models that we construct using a specific OI-EOS described in Sec.~\ref{sec:eos} for several different central energy densities. Further, we also construct NS models, where we substitute the OI-EOS with an EOS with a constant speed of sound for a higher-density region as explained in the same section.

\subsection{Shear Modes}
\label{sec:shear_modes}

Shear modes are present in the spectrum of a NS when the crust has crystallized and supports shear stresses; mathematically, this is reflected by a non-zero shear modulus $\mu$. We have detailed the different shear moduli considered in this study in Sec.~\ref{sec:shearmodulus}.

\begin{figure}
    \centering
    \includegraphics[width=1.0\linewidth]{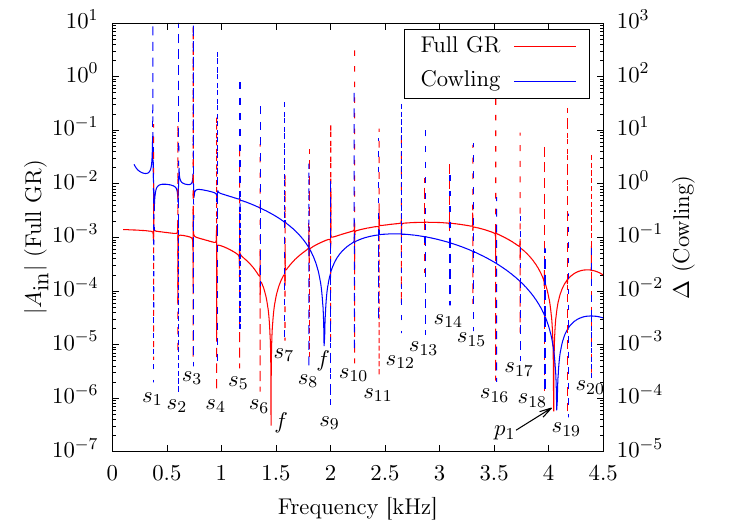}
    \caption{Spectrum of the $l=2$ oscillation modes excited in the NS model with $\epsilon_c = 0.536 \times 10^{15}\textrm{g}/\textrm{cm}^3$. The red line shows the full general relativistic result, while the blue line is the Cowling result; the spikes (which are plotted using dashed lines for clarity as they heavily overlap) indicate the shear modes of which the first 20 are clearly visible. It is apparent that the shear modes are only marginally impacted by the Cowling approximation. Additionally, the frequency of the $f$-mode ($1.455\,$kHz in full GR) is considerably shifted by the Cowling approximation by 33\% to $1.941\,$kHz; the $p_1$-mode is shifted from $4.046\,$kHz to $4.075\,$kHz.}
    \label{fig:s_modes}
\end{figure}

\begin{table}
    \centering
    \caption{Frequencies of the first six $s$-modes with $l =2$ excited in the NS model with $\epsilon_c = 0.536 \times 10^{15}\textrm{g}/\textrm{cm}^3$. The frequencies in the Cowling approximation and full General Relativity coincide very well; the (magnitude of the) relative deviation of the Cowling frequencies from those in full General Relativity is shown in the last column. All frequencies are in the unit of kHz.}
    \begin{tabular*}{80mm}{@{\extracolsep{\fill}}c|ccc}
    \hline
    \hline
       mode  & Cowling & Full GR & rel. err. \\
    \hline
        $s_1$ & 0.3817 & 0.3786 & 0.8\% \\
        $s_2$ & 0.6052 & 0.6020 & 0.5\% \\
        $s_3$ & 0.7456 & 0.7409 & 0.6\% \\
        $s_4$ & 0.9619 & 0.9570 & 0.5\% \\
        $s_5$ & 1.1712 & 1.1666 & 0.4\% \\
        $s_6$ & 1.3584 & 1.3556 & 0.2\% \\
    \hline
    \hline
    \end{tabular*}
    \label{tab:a000_0.536}
\end{table}

\begin{table}
    \centering
    \caption{Same as Table~\ref{tab:a000_0.536} but for the NS model with $\epsilon_c = 1.626 \times 10^{15}\textrm{g}/\textrm{cm}^3$ and $c_s^2 = \alpha = 1.0$ for $\epsilon \geq \epsilon_t$. The eigenfunctions of the first four $s$-modes are shown in Figs.~\ref{fig:s_modes_ef_V} and \ref{fig:s_modes_ef_W}. All frequencies are in the unit of kHz.}
    \begin{tabular*}{80mm}{@{\extracolsep{\fill}}c|ccc}
    \hline
    \hline
      mode  & Cowling & Full GR & rel. err. \\
    \hline
        $s_1$ & 1.4241 & 1.4105 & 1.0\% \\
        $s_2$ & 2.3354 & 2.3188 & 0.7\% \\
        $s_3$ & 3.0247 & 3.0116 & 0.4\% \\
        $s_4$ & 3.6367 & 3.6243 & 0.3\% \\
        $s_5$ & 4.4687 & 4.4648 & 0.1\% \\
        $s_6$ & 5.2418 & 5.2387 & 0.1\% \\
    \hline
    \hline
    \end{tabular*}
    \label{tab:a100_1.62}
\end{table}

In Fig.~\ref{fig:s_modes}, we show a comparison of the spectra for the $l=2$ modes excited in the NS model constructed with the original OI-EOS with a central energy density of $\epsilon_c = 0.536 \times 10^{15}\textrm{g}/\textrm{cm}^3$. 
In this figure, the vertical axes, $|A_{\rm in}|$ and $\Delta$, denote some kind of error functions given in Appendix \ref{app:constrain_damping} and in Ref.~\cite{2023PhRvD.107l3025S}. One can determine the eigenfrequencies with these quantities, i.e., the frequencies with $A_{\rm in}=0$ in the full general relativistic framework (more precisely, when evaluating $A_{\rm in}$ for real-valued frequencies only, sharp minima of this function are relevant) and with $\Delta=0$ for the Cowling approximation correspond to the eigenfrequencies of the stellar model.
As mentioned before, the Cowling approximation impacts the $f$- and $p_1$-modes as is visible from the discrepancy between the red (full GR) and blue (Cowling) lines. On the other hand, we find that the $s$-modes are largely unaffected by the Cowling approximation, which is displayed with dashed lines for clarity (in this way, the strongly overlapping spikes of both curves are visible). We also list the frequencies of the first six $s$-modes in Table~\ref{tab:a000_0.536}; the relative difference between the two formalisms remains only below $1.0\,\%$, which may be attributed to numerical accuracy as well as marginal differences in the implementations of the two perturbation codes.

\begin{figure}
    \centering
    \includegraphics[width=1\linewidth]{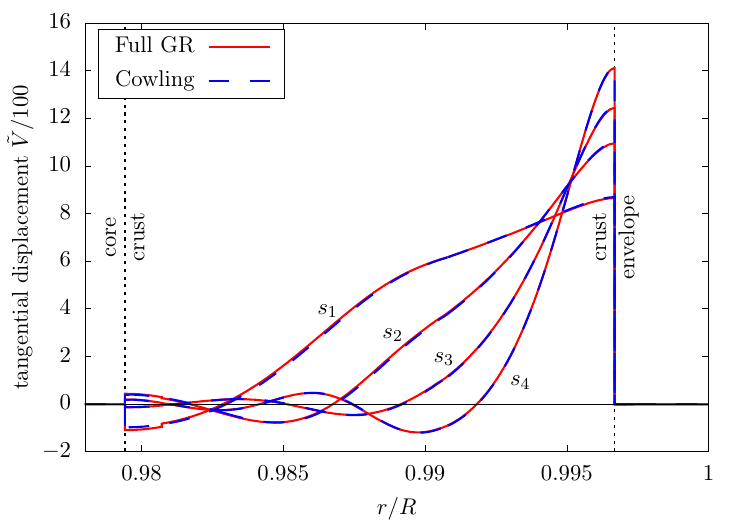}
    \caption{Comparison of the eigenfunctions of the tangential displacement, $\tilde{V}/100$, of the first four $s$-modes between full GR (red solid lines) and the Cowling approximation (blue dashed lines). The NS model has $\epsilon_c = 1.626 \times 10^{15}\textrm{g}/\textrm{cm}^3$ and $c_s^2 = \alpha = 1.0$ for $\epsilon \geq \epsilon_t$. The eigenfunctions are non-zero only in the region where the crust is elastic and it is easy to see that the mode $s_n$ exhibits $n$ zeroes within the crust. It is apparent that the eigenfunctions of the shear modes are virtually unaffected by the Cowling approximation.}
    \label{fig:s_modes_ef_V}
\end{figure}

\begin{figure}
    \centering
    \includegraphics[width=1\linewidth]{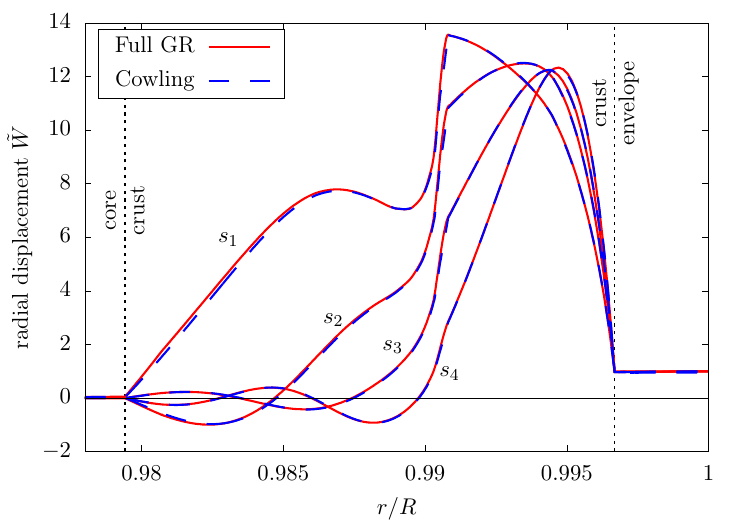}
    \caption{Same as Fig.~\ref{fig:s_modes_ef_V} but for the radial displacement, $\tilde{W}$; again, the eigenfunctions are significantly non-zero only within the crust. Furthermore, comparing the amplitudes shown in Fig.~\ref{fig:s_modes_ef_V} (note that the $y$-axis contains the factor 100) and the present graph, it is clear that $s$-modes exhibit predominantly tangential motion. Again, the eigenfunctions are only marginally affected by the Cowling approximation.}
    \label{fig:s_modes_ef_W}
\end{figure}

We observe that not only the frequencies of the $s$-modes but also their eigenfunctions are essentially unaffected by the Cowling approximation. For the first four $s$-modes with $l=2$ excited in the NS model with $\epsilon_c = 1.626 \times 10^{15}\textrm{g}/\textrm{cm}^3$ and $c_s^2 = \alpha = 1.0$ for $\epsilon \geq \epsilon_t$, we show the eigenfunctions of the tangential displacement, $\tilde{V}$, in  Fig.~\ref{fig:s_modes_ef_V} and those of the radial displacement, $\tilde{W}$, in Fig.~\ref{fig:s_modes_ef_W}, where the results from the code in full General Relativity are plotted with red solid lines, while those from the Cowling code are plotted with blue dashed lines. (The crust of this particular model has a thickness of roughly only 2\% of the star's radius; this may appear thin but is no issue: The crustal thickness decreases considerably when moving close to the maximum mass model \cite{2017MNRAS.470.4397S}.) The overlap of the eigenfunctions of the tangential displacement is nearly impeccable, while those of the radial displacement show marginal discrepancies. We believe that these marginal differences stem from the different implementations of the code and are merely of a numerical nature.

\subsubsection{Damping times of shear modes}

Shear modes couple very weakly to the spacetime; therefore, they are very weak emitters of gravitational waves and their damping time is very large. The imaginary part of the complex-valued frequency $\omega$ is, hence, very difficult to determine due to numerical truncation errors when iterating for the zero of $A_\textrm{in}(\omega)$ in the complex plane. Hence, we resort to the two methods discussed in Sec.~\ref{sec:damp_times} in order to determine the damping times of shear modes.

\begin{table}
    \centering
    \caption{Data for the $s_1$-mode of the NS model with $\epsilon_c = 0.696 \times 10^{15}\textrm{g}/\textrm{cm}^3$ and $\alpha = 0.6$ for the method detailed in Appendix~\ref{app:constrain_damping}. Both poles and zeroes have been determined with high accuracy; the background model is interpolated on more than 60000 grid points. The value for $\alpha_n$ is found by locating the minimum of the Breit-Wigner resonance. Applying the estimates using Eqs.~\eqref{eq:est1} and \eqref{eq:est_beta_n_p} yields a damping time of that mode of $\tau \approx 3.511 \times 10^6\,$s.}
    \begin{tabular}{c|c}
    \hline
    \hline
         & Frequency in kHz \\
    \hline
        $P_1/2\pi$ & 0.843561278872844 \\
        $Z_1/2\pi$ & 0.843561278918188 \\
        $\alpha_n/2\pi$ & 0.843561278918193 \\
        $P_2/2\pi$ & 0.843561278963515 \\
        $Z_2/2\pi$ & 0.843562814080813 \\
    \hline
    \hline
    \end{tabular}
    \label{tab:a060_0.696_pz}
\end{table}

\begin{figure}
    \centering
    \includegraphics[width=1.0\linewidth]{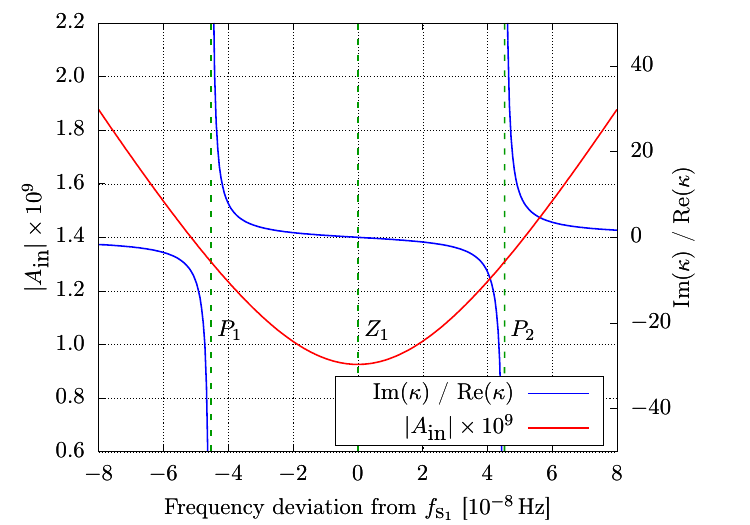}
    \caption{The Breit-Wigner resonance of the $s_1$-mode. The $x$-axis shows the frequencies relative to the minimum of the Breit-Wigner resonance at $f_{\textrm{s}_1} = 0.843561278918194\,$Hz.  The amplitude $|A_\textrm{in}|$ is shown in red; in blue, the phase of the ratio of the amplitudes is shown. It's clearly visible that this curve exhibits two poles $P_1$ and $P_2$ and a zero $Z_1$ (shown with green dashed lines); the latter sits visually at the same location as the minimum of the Breit-Wigner resonance but is in fact $5 \times 10^{-12}\,$Hz smaller. The second zero $Z_2$ is located roughly $1.5\,$mHz to the right.}
    \label{fig:s1_breit_wigner}
\end{figure}

First, we apply the method based on the ratio of the wave amplitudes exemplarily to the $s_1$-mode of the NS model with $\epsilon_c = 0.696 \times 10^{15}\,\textrm{g}/\textrm{cm}^3$ and $\alpha = 0.6$. Figure~\ref{fig:s1_breit_wigner} shows the Breit-Wigner resonance (i.e., the amplitude $|A_\textrm{in}|$) in red, and the phase of the ratio of the amplitudes, ${\rm Im}(\kappa)/{\rm Re}(\kappa)$, in blue. 
In order to precisely determine the poles and zeroes in ${\rm Im}(\kappa)/{\rm Re}(\kappa)$, which we list in Table~\ref{tab:a060_0.696_pz}, a very high frequency resolution of $\Delta\omega \approx 2\pi \times 10^{-9}\,$Hz is required. Quoting frequencies to 15 significant digits is physically, obviously, meaningless; even a different interpolation method for the tabulated EOS or a different background resolution will most likely impact those values. However, in our analysis, not the absolute values but the differences between those frequencies are relevant and these do require this large number of significant digits.
Even though we can determine both poles and both zeroes, the direct evaluation of Eq.~\eqref{eq:const_alpha} yields unreasonable results as the value for $\alpha_n$ comes out smaller than $P_1$ (but it should lie between $Z_1$ and $P_2$ as shown in Table~\ref{tab:a060_0.696_pz}). This is due to the fact that the zero $Z_2$ $(> Z_1)$ is comparably far away from the actual eigenfrequency and the assumption of linearity of the amplitudes $A_\textrm{in}$ and $A_\textrm{out}$, on which the method is built, is violated. However, we may still take the minimum of the Breit-Wigner resonance on the real axis, which is located at $f_{\textrm{s}_1} = \alpha_n / 2\pi \approx 843.6\,$Hz.

Trying to avoid using $Z_2$ leaves us with three options. First, we can simply use Eq.~\eqref{eq:beta_n_poles} which requires the knowledge of the triple $(P_1, P_2, \alpha_n)$ and yields $\tau \approx 3.5106 \times 10^6\,$s. Second, with only the two poles $P_1$ and $P_2$ (not requiring $\alpha_n$), we can apply Eq.~\eqref{eq:est_beta_n_p} and find the interval $3.5106 \times 10^6\,\textrm{s} < \tau \le 4.9647 \times 10^6\,\textrm{s}$. Third, we can use the zero $Z_1$ and the pole $P_2$ adjacent to the eigenfrequency and plug them into Eq.~\eqref{eq:est1} which yields $2.9088 \times 10^6\,\textrm{s} \le \tau < 3.5113 \times 10^6\,\textrm{s}$. The two intervals have a small overlap region which also contains the value for $\tau$ obtained using Eq.~\eqref{eq:beta_n_poles}; we deduce $\tau \approx 3.511 \times 10^6\,$s for the $s_1$-mode which is roughly two months. These values are obtained using more than 60000 grid points inside the star. With a 10-fold lower number of grid points, the absolute values of the poles and zeroes marginally shift but the estimate for the damping time remains virtually the same.

The presented method requires some manual double-checking of the reliability of the results (in particular for very slowly damped modes) and also can provide only a limited number of significant digits due to the close proximity of the poles and zeroes; however, an advantage is that root-finding methods can be used in order to locate the poles and zeroes and such methods usually converge fast and with high accuracy. A direct iteration in the complex plane for the eigenmode would in principle be possible as well, but the solution of a minimization problem is computationally more expensive, and in our experience, the minimizer requires meticulous fine-tuning in the initial values. Further, the naturally finite precision of numerical calculations usually prevents minima from being determined with the same accuracy as roots~\cite{Press2007}. Hence, the presented method provides a useful alternative for estimating damping times of slowly damped modes.

\begin{table}
    \centering
    \caption{Damping times of first six $s$-modes (for $l=2$) of the NS model with $\epsilon_c = 0.696 \times 10^{15}\textrm{g}/\textrm{cm}^3$ and $\alpha = 0.6$. We find that the damping time of the $s_1$-mode is the largest and the next five overtones have damping times that are a few orders of magnitude smaller. PAR denotes the damping times estimated via the phase of the amplitude ratio, while N/SQF and R/RQF are estimates employing variants of the quadrupole formula.}
    \begin{tabular}{c|ccc}
    \hline
    \hline
       Mode  & \multicolumn{3}{c}{Damping time $[10^3\,\textrm{s}]$} \\
         & PAR & N/SQF & R/RQF \\
    \hline
        $s_1$ & 3511\phantom{.000}            & 2205\phantom{.000} & 3533\phantom{.000} \\
        $s_2$ & \hphantom{0}169\phantom{.000} & \hphantom{0}261\phantom{.000} & \hphantom{0}409\phantom{.000} \\
        $s_3$ & \hphantom{000}0.087           & \hphantom{000}0.141           & \hphantom{000}0.216           \\
        $s_4$ & \hphantom{00}13.4\phantom{00} & \hphantom{00}34.1\phantom{00} & \hphantom{00}51.6\phantom{00} \\
        $s_5$ & \hphantom{00}36.0\phantom{00} & \hphantom{0}145\phantom{.000} & \hphantom{0}212\phantom{.000} \\
        $s_6$ & \hphantom{00}38.6\phantom{00} & \hphantom{0}152\phantom{.000} & \hphantom{0}213\phantom{.000} \\
    \hline
    \hline
    \end{tabular}
    \label{tab:a060_0.696_damptimes}
\end{table}

After having laid out in detail the procedure to determine the damping time of a shear ($s_1$-) mode by analysing its the features of the amplitude ratio and by employing two variants of the quadrupole formula, we continue estimating the damping times of its overtones of the same NS model. The course of action is the same for the next five overtones and we show their damping times in Table~\ref{tab:a060_0.696_damptimes}. We observe that the damping times of the overtones first decrease with increasing order up to the third shear mode ($s_3$-mode); subsequently, for higher overtones, the damping times increase again. The damping time of the $s_3$-mode is, somewhat surprisingly, so short that we are able to iterate for the complex frequency $\omega$ directly in the complex plane: We find $\tau_{s_3} = 87.566\,$s while analyzing the amplitude ratio, in very good agreement, yields $\tau_{s_3} = 87.544\,$s. To confirm this behavior of the damping times, we have calculated them using different resolutions of the background model and find the same results. Furthermore, we note that the two variants of the quadrupole formula systematically underestimate the damping time and the relativistic modification R/RQF results in larger deviations than the Newtonian variant N/SQF. While we believe the estimate via the phase of the amplitude ratio to be accurate to percent level, both quadrupole formula estimates yield at least an order of magnitude estimate for the damping time of the $s$-modes.

\subsubsection{Characteristics of $s$-modes for different values of $l$}

\begin{figure}
    \centering
    \includegraphics[width=1.0\linewidth]{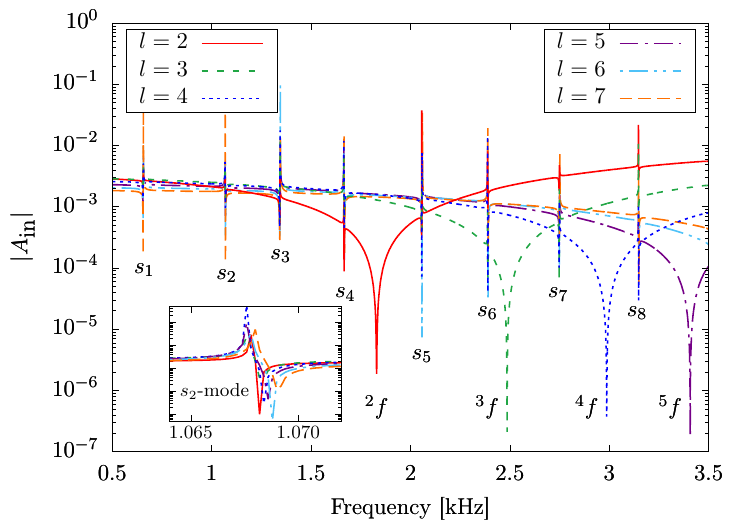}
    \caption{Spectrum of the NS model with $\epsilon_c = 0.907 \times 10^{15}\textrm{g}/\textrm{cm}^3$ for different values of $l$. While the $f$-mode frequency changes considerably, the $s$-mode frequency is only marginally affected. A list of some $s$-mode frequencies is shown in Table~\ref{tab:l-depend}. The overlap of the six curves at the $s$-mode frequencies is not easy to see in the main graph; this is better visible in the inset at the bottom left which shows a magnification around the spikes for the $s_2$-mode.}
    \label{fig:l-depend}
\end{figure}

Having established a way of determining (or at least estimating) the damping times of $s$-modes, we will now, for completeness, investigate the $s$-modes for higher polar order $l>2$. It is well-known that the frequencies of the $f$- and $p$-modes depend strongly on $l$; their frequencies increase with increasing $l$. For the $s$-modes, however, we find that the frequencies are nearly independent of $l$; this has, to the best of our knowledge, been pointed out only once by Vavoulidis \etal~\cite{2008MNRAS.384.1711V}. We show the spectra for $2 \le l \le 7$ of the NS model with $\epsilon_c = 0.907 \times 10^{15}\textrm{g}/\textrm{cm}^3$ in Fig.~\ref{fig:l-depend}. The first eight $s$-modes are clearly visible; their frequency is virtually insensitive to the value of $l$ (in contrast to the $f$-mode frequency). We show the frequencies exemplarily for three of the $s$-modes, i.e., $s_1$-, $s_2$-, and $s_5$-modes, in Table~\ref{tab:l-depend}; they are not identical for different values of $l$ but vary only by tiny amounts. Except for the $s_1$-mode, whose frequency shows an opposite behavior up to $l=5$, the frequencies very slowly increase with increasing value of $l$. 

The damping times, however, change drastically with increasing value of $l$. Making use of the features of the associated amplitude ratios, we are able to estimate the damping times of the $s$-modes for $l=2$ and $l=3$; we show those values along with estimates from the quadrupole formulas in Table~\ref{tab:l-depend_damp}. For higher values of $l$ (i.e., $l \ge 4$), the damping times are so large that a frequency resolution of $\Delta \omega = 4\pi \times 10^{-14}$ is not sufficient to resolve the different poles and zeroes. We can, therefore, only provide the lower bound $\tau > 8000 \times 10^6\,\textrm{s} \approx 250\,\textrm{yrs}$ for the damping times of higher order $s$-modes. The damping times obtained by means of the quadrupole formulas, again, systematically underestimate the values obtained by analysing the phase of the amplitude ratio; however, in contrast to the results shown in Tab.~\ref{tab:a060_0.696_damptimes}, the values from the relativistic modification R/RQF deviate less from the expected value than those from the N/SQF expression.

Looking at Table~\ref{tab:l-depend_damp}, we notice that the damping times of the $s$-modes with $l=3$ are very roughly three orders of magnitude larger than those for $l=2$ (our results indicate that only a slightly larger frequency resolution could reveal the features for the $l=3$ $s_1$-mode, suggesting that a similar factor also holds for this mode). Not being able to resolve the features for the $l=4$ $s_5$-mode, we expect a similar increase in damping times for higher values of $l$, too. We note that the eigenfunctions of the $s$-modes of different $l$ are qualitatively quite similar; this is particularly true within the solid part of the crust where those functions are significantly different from zero and, hence, provide their main contribution to the integrals in the expressions \eqref{eq:E_kin_R} and \eqref{eq:sqf}. However, the amplification factors vary slightly between different modes and also the eigenfunctions of the perturbed densities (both rest-mass and energy) display some larger mismatches. If those eigenfunctions were ``identical'' (up to certain amplification factors), the damping time of an $l>2$ $s$-mode could be calculated from the damping time of the $l=2$ $s$-mode. Our current data set is not large enough to provide a reliable estimate for these amplification factors at this time.

\begin{table}
    \centering
    \caption{Frequencies of three $s$-modes for several values of $l$ of the NS model with $\epsilon_c = 0.907 \times 10^{15}\textrm{g}/\textrm{cm}^3$. The frequency is only marginally affected by the value of $l$. All frequencies are in the unit of Hz.
}
    \begin{tabular*}{80mm}{@{\extracolsep{\fill}}c|ccc}
    \hline
    \hline
      $l$  & $s_1$ & $s_2$ & $s_5$ \\
    \hline
        2 & 654.887 & 1068.20 & 2056.48 \\
        3 & 654.811 & 1068.29 & 2056.72 \\
        4 & 654.745 & 1068.41 & 2057.06 \\
        5 & 654.716 & 1068.58 & 2057.49 \\
        6 & 654.757 & 1068.80 & 2058.01 \\
        7 & 654.904 & 1069.08 & 2058.61 \\
    \hline
    \hline
    \end{tabular*}
    \label{tab:l-depend}
\end{table}

\begin{table}
    \centering
    \caption{Damping times of the $s$-modes reported in Table~\ref{tab:l-depend} in multiples of  $10^6\,$s; they increase quickly with increasing $l$. We are unable to determine numerically the damping times of $s$-modes for $l \ge 4$ due to finite resolution, but can only provide a lower bound of $\tau \gtrsim 10^{10}\,$s for these. PAR denotes the damping times estimated via the phase of the amplitude ratio, while N/SQF and R/RQF are estimates employing variants of the quadrupole formula.}
    \begin{tabular*}{80mm}{@{\extracolsep{\fill}}cc|ccc}
    \hline
    \hline
       &  & \multicolumn{3}{c}{Damping time $[10^6\,\textrm{s}]$} \\
      $l$ & Method & $s_1$ & $s_2$ & $s_5$ \\
    \hline
        \multirow{3}{*}{2} & PAR & 15.2 & 2.29 & 0.0099 \\
         & N/SQF & 11.0 & 3.20 & 0.029\hphantom{0} \\
         & R/RQF & 14.8 & 4.22 & 0.037\hphantom{0} \\
    \hline
        \multirow{3}{*}{3} & PAR & $>8000$ & $>8000$ & 6.46 \\
         & N/SQF & 30000 & 4836 & 10.7 \\
         & R/RQF & 40400 & 6399 & 14.1 \\
    \hline
    \hline
    \end{tabular*}
    \label{tab:l-depend_damp}
\end{table}

\subsection{Interface Modes}

The discontinuities of the shear modulus $\mu$ at the boundaries of the elastic crust give rise to $i$-modes. In contrast to other mode classes (like $p$-, $s$-, or $w$-modes), these modes do not have overtones; instead, there is one $i$-mode per discontinuity (these modes may also arise due to discontinuities in the density but we do not consider those in this study). Hence, we expect to find two $i$-modes; one associated with the crust-core interface and the other related to the crust-envelope interface. The frequencies of the $i$-modes depend on several factors such as depth inside the star as measured from the surface, the magnitude of the discontinuity, or which quantity actually is discontinuous, but is usually of the order of a few tens of Hz \cite{2023PhRvD.107l3025S}. It is, hence, not easy to predict the precise $i$-mode frequencies or, the other way around, to guess from the frequency which discontinuity they belong to: We simply label the $i$-modes with ascending numbers starting from the $i$-mode with the highest frequency.

\begin{figure}
    \centering
    \includegraphics[width=1.0\linewidth]{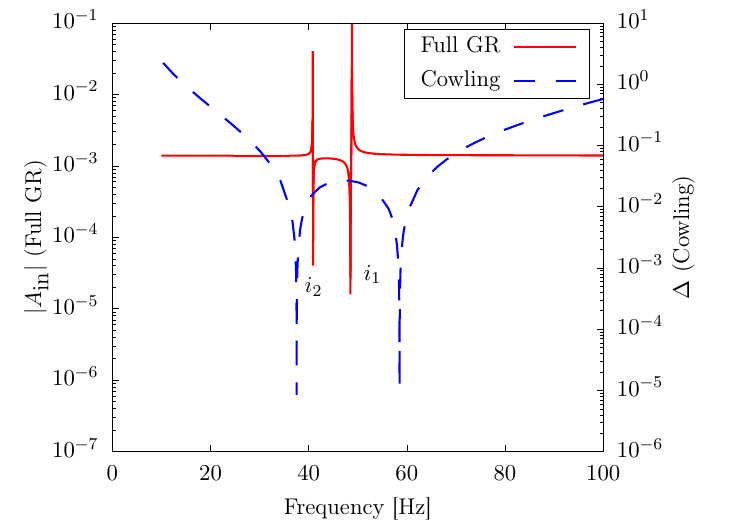}
    \caption{Spectrum of the $l=2$ oscillation modes excited in the NS model with $\epsilon_c = 0.536 \times 10^{15}\textrm{g}/\textrm{cm}^3$; this is the continuation of Fig.~\ref{fig:s_modes} to low frequencies. The red line shows the full general relativistic result, while the blue line is obtained in the Cowling approximation; for both curves, their sharp minima indicate interface modes. We observe a discrepancy in the frequencies of the $i$-modes between the full GR and the Cowling approximation. The frequencies of the modes can be found in Table~\ref{tab:a000_0.536_imodes}.}
    \label{fig:spectrum_i_modes}
\end{figure}

In Fig.~\ref{fig:spectrum_i_modes}, we show the low-frequency domain of the spectrum for the NS model with $\epsilon_c = 0.536 \times 10^{15}\textrm{g}/\textrm{cm}^3$. Contrary to the shear modes (cf.~Sec.~\ref{sec:shear_modes}), whose frequencies displayed perfect agreement between the Cowling results and those in full General Relativity, we observe differing frequencies for the $i$-modes. We explicitly show the frequencies of the two previously considered models in Tables~\ref{tab:a000_0.536_imodes} and \ref{tab:a100_1.62_imodes}; other NS models show comparable behavior.

\begin{table}
    \centering
    \caption{Frequencies of the $i_1$- and $i_2$-modes excited in the NS model with the original OI-EOS, assuming $\epsilon_c = 0.536 \times 10^{15}\textrm{g}/\textrm{cm}^3$. The frequencies in the Cowling approximation and full General Relativity are of the same order of magnitude; the (magnitude of the) relative deviation of the Cowling frequencies from those in full General Relativity is shown in the last column. All frequencies are in the unit of Hz.}
    \begin{tabular*}{80mm}{@{\extracolsep{\fill}}c|ccc}
    \hline
    \hline
        mode & Cowling & Full GR & rel. err. \\
    \hline
        $i_1$ & 58.23 & 48.51 & 20\% \\
        $i_2$ & 37.56 & 40.91 & 8.2\% \\
    \hline
    \hline
    \end{tabular*}
    \label{tab:a000_0.536_imodes}
\end{table}

\begin{figure}
    \centering
    \includegraphics[width=1\linewidth]{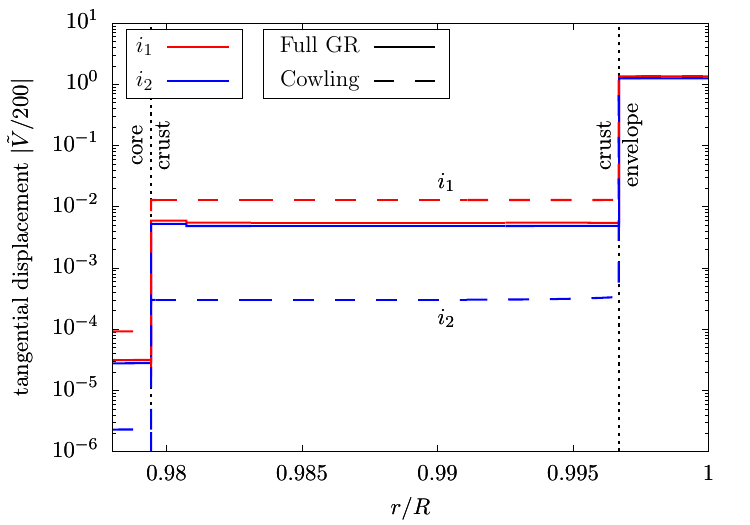}
    \caption{Comparison of the eigenfunctions of the tangential displacement $\tilde{V}$ of the two interface modes of the NS model with $\epsilon_c = 1.626 \times 10^{15}\,\textrm{g}/\textrm{cm}^3$ and $\alpha = 1.0$; We show the $i_1$-mode in red and the $i_2$-mode in blue; full GR and Cowling results are distinguished by solid and dashed lines, respectively. Besides the difference in frequency (cf.~Table~\ref{tab:a100_1.62_imodes}) between the full GR solution and the Cowling approximation, we also observe a difference in the eigenfunctions: for the $i_1$-mode the discontinuity is stronger in full GR, while it is larger in the Cowling approximation for the $i_2$-mode.}
    \label{fig:i_modes_ef_V}
\end{figure}

\begin{figure}
    \centering
    \includegraphics[width=1\linewidth]{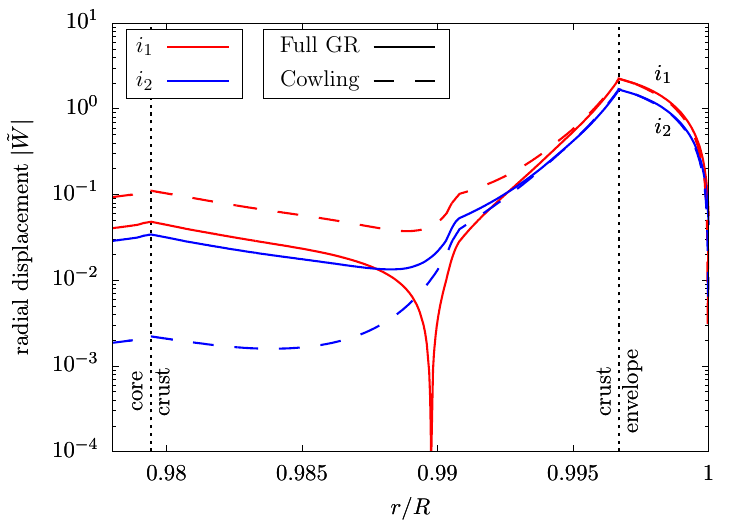}
    \caption{Same as Fig.~\ref{fig:i_modes_ef_V} but for the radial displacement $\tilde{W}$. The radial displacement of the interface modes does not differ much in the envelope, however, inside the solid crust the eigenfunctions start to deviate from each other. Crucially, the eigenfunction of the $i_1$-mode exhibits a node in the solid crust in the full GR case, while it does not perform a sign change in the Cowling approximation.}
    \label{fig:i_modes_ef_W}
\end{figure}

Further, in Figs.~\ref{fig:i_modes_ef_V} and \ref{fig:i_modes_ef_W}, we show the eigenfunctions of the $i$-modes listed in Table \ref{tab:a100_1.62_imodes}; note that, as has been the case for the $s$-modes, the tangential displacement $\tilde{V}$ is a couple of orders of magnitude larger than the radial displacement $\tilde{W}$. The radial displacement shows the largest amplitude at the crust-envelope interface for both $i$-modes.

The eigenfunctions of the tangential displacement $\tilde{V}$ resemble a step function with discontinuities at the two boundaries of the crust. In contrast to the shear modes, cf.~Fig.~\ref{fig:s_modes_ef_V}, the interface modes are non-zero also outside the crust; we have normalized the eigenfunction such that they have the same amplitude in the envelope of the star. Besides the difference in frequency (the frequencies of the modes can be found in Table~\ref{tab:a100_1.62_imodes}) between the full GR solution and the Cowling approximation, we also observe that the eigenfunctions differ from each other; the discontinuities at the interfaces are larger in full General Relativity for the $i_1$-mode, but they are larger in the Cowling approximation for the $i_2$-mode. Qualitatively, though, the eigenfunctions are similar.

Figure~\ref{fig:i_modes_ef_W} shows the eigenfunctions of the radial displacement $\tilde{W}$ of the $i$-modes. We intentionally give them a slightly different amplitude between the $i_1$- and $i_2$-modes in order to be able to follow their curves deeper into the star. A characteristic feature of the $i$-modes is that their eigenfunctions exhibit cusps at the interfaces; these are clearly visible. The radial displacement is clearly non-zero both in the core and the envelope of the star. In the envelope, there is virtually no difference between the eigenfunctions in full GR and the Cowling approximation; however, when going deeper inside the star, the eigenfunctions start to deviate from each other. The eigenfunctions of the $i_2$-mode remain qualitatively fairly similar, while the eigenfunction of the $i_1$-mode exhibits a node inside the solid crust in full GR but does not do so in the Cowling approximation.

The discrepancy of the frequencies as well as eigenfunctions of the $i$-modes is somewhat unexpected. As they are modes that couple only very weakly to the spacetime and hardly radiate gravitational waves, we would expect them to be virtually unaffected by the Cowling approximation in the same way that we have observed it for the $s$-modes (cf.~Sec.~\ref{sec:shear_modes}). 
In order to resolve this observation, we have first compared the background solution of our two independent TOV solvers; they yield nearly identical NS models whose radii, at which the surface and the crustal boundaries are located, differ only on the order of centimeters. While this may have an impact on the resulting frequencies, we are not convinced that this can explain the observed frequency discrepancies of several percent.

We also investigate the damping time of the $i$-modes. For both interface modes of the NS model with $\epsilon_c = 1.626 \times 10^{15}\textrm{g}/\textrm{cm}^3$ and $\alpha = 1.0$ we are unable to resolve the individual poles and zeroes that are required for the method detailed in App.~\ref{app:constrain_damping}. To be precise, at a frequency resolution of $\Delta \omega = 2\pi \times 10^{-9}\,$Hz the necessary features cannot be resolved; at a 10-fold higher frequency resolution $\Delta \omega = 2\pi \times 10^{-10}\,$Hz numerical inaccuracies are visible in the graph which prevent an accurate determination of the phase of the  amplitude ratio. We can, therefore, only conclude that the damping time $\tau$ of both interface modes is $\tau > 3.18 \times 10^{11}\,\textrm{ms} \approx 10\,\textrm{yrs}$ (since $\beta_n < \Delta\omega/2 = \pi \times 10^{-9}\,\textrm{Hz}$ according to our analysis in App.~\ref{app:constrain_damping}).

We did not expect a short damping time for the $i$-modes, however, our finding of a very long damping time also supports the idea that the Cowling approximation should not have a large impact on these modes. Having compared our calculations in full GR and the Cowling approximation in detail, we can, at the moment, only report our finding that we find slightly differing results. The important message that we take from this investigation is that even minute differences in the background models that seem to be negligible may result in differences in the results that are larger than expected (without being numerical instabilities); in particular, $i$-modes seem to be sensitive to such small discrepancies.

\begin{table}
    \centering
    \caption{Same as Table~\ref{tab:a000_0.536_imodes} but for the NS model with $\epsilon_c = 1.626 \times 10^{15}\textrm{g}/\textrm{cm}^3$ and $c_s^2 = \alpha = 1.0$ for $\epsilon \geq \epsilon_t$. All frequencies are in the unit of Hz. The corresponding eigenfunctions are shown in Figs.~\ref{fig:i_modes_ef_V} and \ref{fig:i_modes_ef_W}.}
    \begin{tabular*}{80mm}{@{\extracolsep{\fill}}c|ccc}
    \hline
    \hline
       mode  & Cowling & Full GR & rel. err. \\
    \hline
        $i_1$ & 25.3 & 22.36 & 13.1\% \\
        $i_2$ & 20.5 & 21.71 & 5.6\% \\
    \hline
    \hline
    \end{tabular*}
    \label{tab:a100_1.62_imodes}
\end{table}

\section{Conclusions and Outlook}
\label{sec:conclusions}

We construct relativistic neutron star models that possess an elastic, albeit unstrained, crust composed of spherical and cylindrical nuclei. On these background models, we calculate shear and interface modes, which are due to the presence of the elastic crust and its boundaries, respectively, and compare their characteristics in full General Relativity as well as the relativistic Cowling approximation (in which spacetime perturbations are neglected). We find that the frequency as well as the eigenfunctions of the shear modes are virtually unaffected by the Cowling approximation and the marginal differences between the obtained frequencies can be attributed to numerical inaccuracies in the respective implementations of the independently developed codes. Additionally, we observe that the frequency of the shear modes is only marginally affected by the polar quantum number $l$. In contrast, we find that the frequencies of the interface modes are considerably by the Cowling approximation; their eigenfunctions differ from each other, too. This result is somewhat unexpected and the discrepancies in the frequencies and eigenfunctions are considerably larger than the marginal differences that we detected between the solutions of our respective TOV solvers. While we do not commit ourselves to stating which of our two numerical codes produces the ``correct'' results, we simply advise caution in the calculation of $i$-modes as even tiny differences in the constructed background model may lead to deviations in the results that are larger than perhaps initially expected.

Further, we have extended a previously published method that utilizes the features of the phase of the amplitude ratio, namely poles and zeroes in its phase, which are associated with each mode, in order to determine the damping times of slowly damped modes. We demonstrate the accuracy of this method and how to apply it on the $f$- and $p$-modes, whose damping times can be accurately determined also by minimizing the error function $|A_\textrm{in}|$ in the complex plane. Sometimes, not all poles or zeroes can be determined or it could be the case that their location is beyond the validity of the assumed linearity approximation. For these cases, where only a subset of the features is accurately known, we show how the damping time of a slowly damped mode can still be accurately calculated or at least be robustly constrained. Employing these extensions, we accurately determine the damping times of shear modes; we find that the first few overtones have considerably shorter damping times than the fundamental shear mode. When increasing the polar order $l$, the damping time of the shear modes quickly increases, too. We also employ the quadrupole formula for the damping times and find that it yields good order-of-magnitude estimates. We also attempt to estimate the damping time of the interface modes; however, these are so long that we can only provide a lower bound of $\approx 10\,$yrs for them.

While the negligible impact of the Cowling approximation on shear modes was expected, this study fills a gap in performing a quantitative analysis on this issue. The present work focused on a few selected NS models and we expect comparable results if we had studied different NS models; further, this study presents how damping times of very slowly damped modes can be estimated. As future advancements of this study, it could be interesting to investigate the sensitivity of the $i$-modes on the background model and to employ different EOSs in order to investigate also the impact of the Cowling approximation on $i$-modes due to density discontinuities (rather than discontinuities in the shear modulus).

\begin{acknowledgments}
We thank the anonymous referee for very constructive comments that helped improve the content of this manuscript.
C.K. warmly thanks the Interdisciplinary Theoretical and Mathematical Science Program (iTHEMS) of RIKEN  for their kind hospitality and support by the iTHEMS working group ``GW-EOS", where this work has been completed. This work is supported in part by Japan Society for the Promotion of Science (JSPS) KAKENHI Grant Numbers JP23K20848 and JP24KF0090, and by FY2023 RIKEN Incentive Research Project.
\end{acknowledgments}

\appendix
\section{Perturbation Equations for the Elastic Crust in full GR}
\label{sec:perteq_elastic_fullgr}

The differential equations governing polar perturbations of the elastic crust in full General Relativity have been derived and presented in Ref.~\cite{2015PhRvD..92f3009K}. However, as there has been a small mistake in the derivation of the equations, we will repeat the perturbation equations here in the correct form; the error in Ref.~\cite{2015PhRvD..92f3009K} was a factor of 2 in the shear modulus $\mu$ that was missing.

We define the radial traction, $T_1:= \delta \Sigma_r^{\ r}$, and the tangential traction, $T_2:= \delta \Sigma_r^{\ \theta}$, as done in Refs.~\cite{2011PhRvD..84j3006P, 2015PhRvD..92f3009K}. Further, we define the perturbation variable $X:= -r^{-l} e^\Phi \Delta p$. We also define the two abbreviations:
\begin{equation}
    Q := r^2 e^{-2\Lambda} \Phi'
    \quad\text{and}\quad
    n := \frac{1}{2} (l+2) (l-1),
\end{equation}
and make liberal use of the TOV equations. The full set of perturbation equations for the elastic crust then is
\begin{align}
    \frac{\partial H_1}{\partial r}
        & = \left[ (\Lambda' - \Phi')
                     - \frac{l+1}{r} \right]
              H_1
    \nonumber\\
        & \quad + \frac{e^{2\Lambda}}{r}
                \left[ H_2 + K - 16 \pi (\epsilon + p) \tilde{V} \right],
        \label{eq:odeH1} \\
    \frac{\partial K}{\partial r}
        & = \frac{1}{r} H_2
            + \frac{n+1}{r}H_1
            + \left( \Phi' - \frac{l+1}{r} \right) K
    \nonumber\\
        & \quad
            - \frac{8 \pi (\epsilon + p) e^{\Lambda} }{r} \tilde{W},
            \label{eq:odeK} \\
    \frac{\partial H_0}{\partial r}
        & = \frac{\partial K}{\partial r}
            -r e^{-2\Phi} \omega^2 H_1
            - \left( \Phi'
            + \frac{l-1}{r} \right) H_0
    \nonumber\\
        & \quad - \left( \Phi'
            + \frac{1}{r}   \right) H_2
            + \frac{l}{r} K
            - \frac{16 \pi}{r} T_2,
            \label{eq:odeH0} \\
    \frac{\partial \tilde{W}}{\partial r}
        & = - \frac{l+1}{r} \tilde{W}
    \nonumber\\
        & \quad + r e^{\Lambda} \left[
                \frac{e^{-\Phi}}{\gamma p} X
                - \frac{l(l+1)}{r^2} \tilde{V}
                + \frac{1}{2} H_2
                + K
            \right],
            \label{eq:odeW} \\
    \frac{\partial \tilde{V}}{\partial r}
        & = \frac{1}{\mu r} T_2
            + \frac{e^{\Lambda}}{r} \tilde{W}
            + \frac{2-l}{r} \tilde{V},
            \label{eq:odeV} \\
    \frac{\partial T_2}{\partial r}
        & = - \frac{1}{2} r e^{2\Lambda} (\epsilon + p) H_0
            + r e^{2\Lambda - \Phi} X
            - \frac{e^{2\Lambda}}{2r} T_1
    \nonumber\\
        & \quad + \left[
                (\Lambda' - \Phi')-\frac{l+1}{r}
              \right] T_2
            \nonumber \\
        & \quad + \left[
                    \frac{2 n \mu e^{2\Lambda}}{r}
                    -e^{2\Lambda-2\Phi} r \omega^2 (\epsilon + p)
                  \right] \tilde{V}
                + e^{\Lambda} p' \tilde{W}.
                \label{eq:odeT2}
\end{align}
In addition to these six ordinary differential equations, we have three
algebraic relations:
\begin{align}
    H_2
        & = H_0 + 32 \pi \mu \tilde{V},
        \label{eq:alg1} \\
    0
        & = 8\pi r^3 e^{-\Phi} X + 8 \pi r T_1
    \nonumber\\
        & \quad 
             - \left( 2M + Q + nr \right) H_0
             - 16 \pi r e^{-2\Lambda} T_2
    \nonumber\\
        & \quad - \left[ (n+1)Q - \omega^2 r^3 e^{-2(\Lambda+\Phi)} \right] H_1
                \nonumber \\
        & \quad + \left[ nr - \omega^2 r^3 e^{-2\Phi}
                     - \frac{e^{2\Lambda}}{r} Q (2M + Q - r) \right] K,
                \label{eq:alg2} \\
    0
        & = \frac{\mu r^2 e^{-\Phi}}{3} X - \frac{1}{4} \gamma p T_1
    \nonumber\\
        & \quad - \mu \gamma p
            \left[ e^{-\Lambda} \tilde{W} - \frac{r^2}{2} K + (n+1) \tilde{V} \right].
                \label{eq:alg3}
\end{align}
The first of these three algebraic relations can be used to determine $H_2$ out of $H_0$ and $\tilde{V}$; the latter two are a linear system that can be solved for $X$ and $T_1$.

The perturbation equations have to be solved in different layers of the star; the individual solutions have to be linked to each other by appropriate junction conditions which stem from the requirement that the intrinsic curvature be continuous across these interfaces~\cite{1990MNRAS.245...82F, 2011PhRvD..84j3006P}. We refer the reader to Appendix C in Ref.~\cite{2015PhRvD..92f3009K} for a general description of the numerical procedure and the relevant boundary conditions. For the sake of completeness, we repeat here the physical motivation of the junction conditions: The metric perturbations $H_0$, $H_1$, and $K$ have to be continuous across all interfaces inside the star (this concerns the interfaces between the perfect fluid and the elastic crust, but also the interface within the elastic crust where the shear modulus is discontinuous); further, the radial displacement encoded in $\tilde{W}$ must be continuous since otherwise a ``gap'' in the NS matter could occur; last, the tangential traction $T_2$ must be continuous as well. As a consequence, the Lagrangian pressure perturbation represented by $X$ is also continuous across these interfaces, however, this condition is redundant.

\section{Constraining damping times of long-lived QNMs}
\label{app:constrain_damping}

Determining the damping time, i.e., the imaginary part of the complex-valued frequency $\omega = \sigma + i / \tau$, of a quasi-normal mode (QNM) poses numerical difficulties for long-lived modes for which $1/ \tau \ll \sigma$. This is because the frequency $\omega$ appears squared in the perturbation equations whereby real and imaginary parts are mixed, resulting in a loss of accuracy of the imaginary part.

This problem has been approached by Andersson {\it et al}.~\cite {2002PhRvD..66j4002A} in their Appendix C by investigating so-called Breit-Wigner resonances associated with a QNM; for our study, we revisit this investigation and complement the previous study by a few more details. An accurate estimate of the damping time of long-lived QNMs remains tricky and at times impossible when the damping time is very large (on the other hand, for such long damping times precise knowledge is astrophysically probably not relevant); but in all cases, the method allows to put a lower bound on the damping time $\tau$.

For simplicity, we keep the notation used in Ref. \cite{2002PhRvD..66j4002A} (however, we capitalize the variables for the poles and zeroes to avoid confusion with $p_1$- and $p_2$-modes) and repeat the most important details: The method is based on the ratio $\kappa(\omega) := A_\textrm{out}(\omega) / A_\textrm{in}(\omega)$ of the amplitudes of the outgoing wave $A_\textrm{out}$ and the ingoing wave $A_\textrm{in}$ at infinity. The relativistic system has a QNM with frequency $\omega_n = \alpha_n + i\beta_n$ if $A_\textrm{in}(\omega_n) = 0$; at the same time, we have $A_\textrm{out}(\overline{\omega}_n) = 0$.

The method relies on the assumption that both $A_\textrm{in}$ and $A_\textrm{out}$ depend approximately linearly on $\omega$ in a vicinity of $\omega_n$ and $\overline{\omega}_n$, respectively, i.e.,
\begin{align}
    A_\textrm{in} & \propto \omega - \omega_n,
    \\
    A_\textrm{out} & \propto \omega - \overline{\omega}_n.
\end{align}
Hence, we can assume that the ratio of both amplitudes can be written as
\begin{align}
    \kappa
     = \frac{A_\textrm{out}}{A_\textrm{in}}
      \approx \gamma \frac{\omega - \overline{\omega}_n}{\omega - \omega_n},
\end{align}
where $\gamma = \gamma_r + i\gamma_i$ is a complex-valued constant.

We will now consider the phase of $\kappa$ and restrict ourselves to real-valued $\omega$ only; we find
\begin{align}
    \frac{\textrm{Im}(\kappa)}{\textrm{Re}(\kappa)}
    & \approx \frac{\gamma_i \left[ \left(\omega - \alpha_n\right)^2 - \beta_n^2\right] + 2\gamma_r \beta_n (\omega - \alpha_n)}
             {\gamma_r \left[ \left(\omega - \alpha_n\right)^2 - \beta_n^2\right] - 2\gamma_i \beta_n (\omega - \alpha_n)}.
    \label{eq:phase_kappa}
\end{align}
One can easily observe that both the numerator as well as denominator of the right-hand side of Eq.~\eqref{eq:phase_kappa} are quadratic in $\omega$; hence, this function has two poles (denoted with $P_1$ and $P_2$) and two zeroes ($Z_1$ and $Z_2$) on the real axis of $\omega$. These can be determined to be
\begin{align}
    P_{1,2}
    & = \alpha_n + \beta_n \left( \mathcal{G}^{-1} \pm \sqrt{1 + \mathcal{G}^{-2}} \right),
    \label{eq:def_poles}
    \\
    Z_{1,2}
    & = \alpha_n - \beta_n \left( \mathcal{G} \pm \sqrt{1 + \mathcal{G}^2}\right),
    \label{eq:def_zeroes}
\end{align}
where we have introduced the abbreviation $\mathcal{G} := \gamma_r / \gamma_i$ and the signs are chosen such that $P_1 < P_2$ and $Z_1 < Z_2$. It is immediately clear that $P_1 < \alpha_n < P_2$ and $Z_1 < \alpha_n < Z_2$.

Andersson {\it et~al}.~\cite{2002PhRvD..66j4002A} have inverted these formulas for the frequency $\alpha_n$ and damping time\footnote{For ease of readability, we call $\beta_n$ the ``damping time'', even though it actually is its reciprocal.} $\beta_n$ of the QNM as well as the ratio $\gamma_r / \gamma_i = \mathcal{G}$:
\begin{align}
    \alpha_n
    & = \frac{P_1 P_2 - Z_1 Z_2}{P_1 + P_2 - Z_1 - Z_2},
    \label{eq:const_alpha}
    \\
    \beta_n
    & = \sqrt{(\alpha_n - P_1) (P_2 - \alpha_n)}
    \label{eq:beta_n_poles}
    \\
    & = \sqrt{(\alpha_n - Z_1) (Z_2 - \alpha_n)},
    \label{eq:beta_n_zeroes}
    \\
    \mathcal{G}
    & = \frac{\beta_n}{(P_1 + P_2)/2 - \alpha_n}
    \label{eq:phase_gamma_p}
    \\
    &  = \frac{\alpha_n - (Z_1 + Z_2)/2}{\beta_n}.
    \label{eq:phase_gamma_z}
\intertext{Additionally, we have}
    \left| \mathcal{G} \right|
    & = \frac{Z_2 - Z_1}{P_2 - P_1}.
\end{align}

Given that the two poles and two zeroes can be precisely determined, the frequency, $\alpha_n$, and damping time, $\beta_n$, of the QNM can be recovered. Having two formulae for $\beta_n$ and $\mathcal{G}$ allows to perform a double-check; this is useful since the method relies on the fact that $\gamma$ is assumed to be constant (which in reality it is not) across the interval spanned by the poles and zeroes. Any non-negligible discrepancy suggests that this assumption is violated. Note that the relations in Eqs.~\eqref{eq:const_alpha} - \eqref{eq:phase_gamma_z} recover the parameters required for the right-hand side of Eq.~\eqref{eq:phase_kappa}; they are, however, only approximations (albeit very good ones) to the true frequency and damping time of the QNM.

We report another fact about the approximant to $\textrm{Im}(\kappa)/\textrm{Re}(\kappa)$, i.e., the right-hand side of Eq.~\eqref{eq:phase_kappa}. Its derivative with respect to $\omega$ is
\begin{align}
    - \frac{2 \beta_n |\gamma|^2 \left((\omega - \alpha_n)^2 + \beta_n^2 \right)}{\left[ \gamma_r \left[ \left(\omega - \alpha_n\right)^2 - \beta_n^2\right] - 2\gamma_i \beta_n (\omega - \alpha_n) \right]^2} < 0,
\end{align}
which is strictly smaller than zero since $\beta_n > 0$ by definition. The blue curve in Fig.~\ref{fig:s1_breit_wigner} has indeed a negative slope everywhere. This fact is useful when implementing a script to determine the poles and zeroes automatically.

We are now going to present further details about this method which allows for estimates of the damping time $\beta_n$ even when not all two poles and two zeroes can reliably be determined (the frequency $\alpha_n$ can be approximated very well by finding the minimum of $|A_\textrm{in}(\omega)|$ on the real axis) or that could help in locating a ``missing'' pole or zero. In general, given the above definitions, it can be shown that the poles and zeroes can appear only in two possible orders, which depend on the sign of $\mathcal{G}$. It can be shown that either
\begin{align}
    Z_1 & < P_1 < \alpha_n < Z_2 < P_2\quad\text{for}\quad \mathcal{G} > 0,
    \label{eq:z1p1_z2p2}
    \\
\intertext{or}
    P_1 & < Z_1 < \alpha_n < P_2 < Z_2\quad\text{for}\quad \mathcal{G} < 0,
\end{align}
i.e., we will find one pole and one zero on either side of the frequency $\alpha_n$.

As a first strategy, we scan only the vicinity of $\alpha_n$. Scanning with sufficient resolution to the left of $\alpha_n$, we would first encounter either the pole $P_1$, implying $\mathcal{G} > 0$, or the zero $Z_1$, which implies $\mathcal{G} < 0$; let's assume we also determine the pole/zero adjacent to $\alpha_n$ on the right. Let us focus first on the case $\mathcal{G} > 0$; we can calculate the difference $Z_2 - P_1$, which is
\begin{align}
    Z_2 - P_1
    & = \left[ \sqrt{1 + \mathcal{G}^2} + \sqrt{ 1 + \mathcal{G}^{-2}} - \mathcal{G} - \mathcal{G}^{-1}\right] \beta_n.
\end{align}
The prefactor of $\beta_n$ (keeping in mind that $\mathcal{G} > 0$) can be shown to be constrained to the half-open interval $[2\sqrt{2} - 2, 1) \approx [0.83, 1)$. 
Hence, the knowledge of $P_1$ and $Z_2$ allows us to constrain the damping time to
\begin{align}
    Z_2 - P_1 < \beta_n & \le \frac{1}{2(\sqrt{2} - 1)} (Z_2 - P_1)
    \\
    & \approx 1.2 (Z_2 - P_1)\quad\text{for}\quad \mathcal{G} > 0.
\end{align}
Similarly, if scanning for poles and zeroes implies $\mathcal{G} < 0$, we find
\begin{align}
    P_2 - Z_1 < \beta_n & \le \frac{1}{2(\sqrt{2} - 1)} (P_2 - Z_1)
    \label{eq:est1}
    \\
    & \approx 1.2 (P_2 - Z_1)\quad\text{for}\quad \mathcal{G} < 0.
\end{align}
Irrespective of the sign of $\mathcal{G}$, the damping time $\beta_n$ can be constrained to roughly $20\%$ accuracy only by determining the difference between pole and zero directly adjacent to $\alpha_n$. Further, this difference (i.e., $Z_2 - P_1$ for $\mathcal{G} > 0$ and $P_2 - Z_1$ for $\mathcal{G} < 0$) may serve as a crude estimate for the damping time $\beta_n$.

A second strategy for estimating $\beta_n$ can be deduced from the differences between the poles, $\Delta P:= P_2 - P_1$, or the zeroes $\Delta Z:= Z_2 - Z_1$. From the poles and zeroes provided by Eqs.~\eqref{eq:def_poles} and \eqref{eq:def_zeroes}, it is easy to see that $\Delta P = 2 \beta_n \sqrt{1 + \mathcal{G}^{-2}}$ and $\Delta Z = 2 \beta_n \sqrt{1 + \mathcal{G}^2}$. We must have either $\Delta P > \Delta Z$, which implies $|\mathcal{G}| < 1$, or $\Delta P \le \Delta Z$, which implies $|\mathcal{G}| \ge 1$. If we are able to determine which of those two, $\Delta P$ or $\Delta Z$, is the smaller one (perhaps because a high resolution scan reveals two poles but only one zero, implying that the second zero is further away), we can again find a constraint on $\beta_n$. For the case that $|\mathcal{G}| \ge 1$ (or $\Delta P \le \Delta Z$), we have
\begin{align}
    \frac{1}{\sqrt{2}} \frac{P_2 - P_1}{2} & \le \beta_n <  \frac{P_2 - P_1}{2},
    \label{eq:est_beta_n_p}
    \\
\intertext{and for the case $|\mathcal{G}| < 1$ or $\Delta Z < \Delta P$, we have}
    \frac{1}{\sqrt{2}} \frac{Z_2 - Z_1}{2} & < \beta_n <  \frac{Z_2 - Z_1}{2}.
    \label{eq:est_beta_n_z}
\end{align}
Further, we can use $\Delta P/2$ or $\Delta Z/2$ (whichever is smaller) as a crude estimate for $\beta_n$.

The two strategies presented offer constraints or estimates for the damping time $\beta_n$ even when only a subset of the two poles and two zeroes can be determined. As a third strategy, such an estimate could also be calculated from Eqs.~\eqref{eq:beta_n_poles} or \eqref{eq:beta_n_zeroes}. These two expressions require, in addition to either the two poles or two zeroes, the knowledge of the frequency $\alpha_n$, which can in principle be determined with high accuracy as the minimum of the Breit-Wigner resonance. However, the two former strategies work without knowing or estimating $\alpha_n$. This can increase their robustness in light of the high resolution $\Delta \omega$ required to separate the poles and zeroes in the vicinity of $\alpha_n$.

This tells us immediately that the frequency resolution $\Delta \omega$ (or more specifically $\Delta \alpha$) with which $\textrm{Im}(\kappa) / \textrm{Re}(\kappa)$ is evaluated on the $x$-axis in the vicinity of $\alpha_n$ is closely linked to an upper limit on $\beta_n$ (and hence a lower limit on the actual damping time $\tau_n$): If a certain frequency resolution $\Delta \omega$ does not suffice to separate the poles or zeroes---meaning that $\Delta P$ or $\Delta Z$ are smaller than $\Delta \omega$---one can safely assume that $\beta_n < \Delta \omega/2$, owing to Eqs.~\eqref{eq:est_beta_n_p} and \eqref{eq:est_beta_n_z}.

\subsection{Testing the method}
\label{app:kappa_test}

As the method detailed above relies on a linearity approximation in the vicinity of a QNM $\omega_n$, we test its validity for a few modes for which the damping time may also be estimated by iteration in the complex plane; these will be the $f$-mode and the first and second $p$-modes (for $l=2$). As an example, we investigate the NS model constructed with the original OI-EOS, whose central energy density is $\epsilon_c = 0.907 \times 10^{15}\textrm{g}/\textrm{cm}^3$; we also show some part of the spectrum of that star including $s$-modes for $2 \le l \le 7$ in Fig.~\ref{fig:l-depend}.

First, we turn to the first two pressure modes, i.e., $p_1$- and $p_2$-modes. We show the corresponding poles and zeroes for those two modes in Tab.~\ref{tab:a000_0.907_pz_p1p2}; note that a large number of significant digits is not meant to be physical accuracy but we need these as the differences between the values are crucial for the method. For both modes, we can easily identify all four frequencies. Further, we can also iterate for the correct complex-valued frequency $\omega$ in the complex plane using a minimizer for $|A_\textrm{in}(\omega)|$; those values are shown in Tabs.~\ref{tab:a000_0.907_pz_p1_result} and \ref{tab:a000_0.907_pz_p2_result} in the row labeled ``complex iteration''. Alternatively, we can employ the values for the poles and zeroes in order to estimate the frequency of the mode by means of Eq.~\eqref{eq:const_alpha}; for the damping time, we can use both Eq.~\eqref{eq:beta_n_poles} and Eq.~\eqref{eq:beta_n_zeroes} for an estimate, i.e., employing either both poles or both zeroes. We show these estimates in these tables in the rows labeled ``estimate~1'' and ``estimate~2''.

\begin{table}
    \centering
    \caption{Data for the $p_1$- and $p_2$-mode (with $l=2$) of the NS model with $\epsilon_c = 0.907 \times 10^{15}\textrm{g}/\textrm{cm}^3$ based on the original OI-EOS.}
    \begin{tabular}{c|cc}
    \hline
    \hline
         & $p_1$-mode & $p_2$-mode \\
         & [kHz]      & [kHz] \\
    \hline
        $P_1/2\pi$ & 5.837937525 & 7.808827336 \\
        $Z_1/2\pi$ & 5.837972177 & 7.808827917 \\
        $P_2/2\pi$ & 5.838001207 & 7.808828221 \\
        $Z_2/2\pi$ & 5.838329949 & 7.808829189 \\
    \hline
    \hline
    \end{tabular}
    \label{tab:a000_0.907_pz_p1p2}
\end{table}

\begin{table}
    \centering
    \caption{Results for the $p_1$-mode of the NS model with $\epsilon_c = 0.907 \times 10^{15}\textrm{g}/\textrm{cm}^3$ using the data from Tab.~\ref{tab:a000_0.907_pz_p1p2}. The first row shows the correct results obtained from the 2-dimensional minimization problem in the complex plane; the second row, labeled ``estimate~1'' shows the estimates obtained using Eqs.~\eqref{eq:const_alpha}, \eqref{eq:beta_n_poles}, and \eqref{eq:phase_gamma_p}; the third row, labelled ``estimate~2'' shows the estimates obtained using Eqs.~\eqref{eq:beta_n_zeroes}, and \eqref{eq:phase_gamma_z}.}
    \begin{tabular}{c|ccc}
    \hline
    \hline
         & Frequency & Damping time & $\mathcal{G}$ \\
         & [kHz]     & [s]         & \\
    \hline
        complex iteration & 5.837974945 & 5.07694 & -5.620 \\
        estimate 1        & 5.837974945 & 5.07693 & -5.621 \\
        estimate 2        &  ---        & 5.07870 & -5.620 \\
    \hline
    \hline
    \end{tabular}
    \label{tab:a000_0.907_pz_p1_result}
\end{table}

\begin{table}
    \centering
    \caption{Same as Table~\ref{tab:a000_0.907_pz_p1_result} but for the $p_2$-mode of the same NS model.}
    \begin{tabular}{c|ccc}
    \hline
    \hline
         & Frequency & Damping time & $\mathcal{G}$ \\
         & [kHz]     & [s]         & \\
    \hline
        complex iteration & 7.808828031 & 438.268 & -1.436 \\
        estimate 1        & 7.808828023 & 431.626 & -1.508 \\
        estimate 2        &  ---        & 453.390 & -1.510 \\
    \hline
    \hline
    \end{tabular}
    \label{tab:a000_0.907_pz_p2_result}
\end{table}

We observe that the frequency can be recovered to very high accuracy in both cases; the damping time can be estimated very well (to five digits) for the $p_1$-mode, while the damping time of the $p_2$-mode results in an error of $2-3\,\%$. This is still a very good estimate given that we solved the complex-valued problem purely on the real axis. With a higher grid resolution inside the star, the accuracy could be improved further.

Additionally, we show the value of the proportionality constant $\mathcal{G}$ in the last column of Tabs.~\ref{tab:a000_0.907_pz_p1_result} and \ref{tab:a000_0.907_pz_p2_result}. We find numerically that its value indeed varies only marginally in a small environment around the complex-valued eigenfrequency $\omega_n$; this justifies the approximation. We also see that its value can be recovered from the poles and zeroes using Eqs.~\eqref{eq:phase_gamma_p} and \eqref{eq:phase_gamma_z}; again, for the $p_2$-mode the value is a bit off.

We now turn to the $f$-mode which can be easily iterated for in the complex plane. On the real axis, we find two poles and one zero in the vicinity of the $f$-mode frequency (which is $1.829626327\,$kHz, i.e., marginally smaller than $Z_2$), but somewhat surprisingly, we are unable to locate the second zero. Scanning the complex plane, we find that $\mathcal{G} \approx 545 > 0$, which implies that $Z_1 < P_1$ (cf. Eq.~\eqref{eq:z1p1_z2p2}). In fact, we should find $Z_1$ roughly at $1.3\,$kHz; however, the linearity approximation no longer holds at this distance which explains why we are not able to locate a (reliable) corresponding $Z_1$.

\begin{table}
    \centering
    \caption{Data for the $f$-mode of the NS model with $\epsilon_c = 0.907 \times 10^{15}\textrm{g}/\textrm{cm}^3$. The zero $Z_1$ for the $f$-mode should be located at $\approx 1.3\,$kHz; however, its non-existence is due to the linearity approximation no longer being valid there.}
    \begin{tabular}{c|c}
    \hline
    \hline
         & $f$-mode \\
         & [kHz]    \\
    \hline
        $Z_1/2\pi$ & ---         \\
        $P_1/2\pi$ & 1.829143927 \\
        $Z_2/2\pi$ & 1.829626828 \\
        $P_2/2\pi$ & 1.830110615 \\
    \hline
    \hline
    \end{tabular}
    \label{tab:a000_0.907_pz_f}
\end{table}

Using the correct value of $\alpha_n$ (e.g., by finding the minimum of the Breit-Wigner resonance for real-values $\omega$), we can employ $P_1$ and $P_2$ in Eq.~\eqref{eq:beta_n_poles} and find $\tau = 329.27942\,$ms which is only marginally different from the correct value $\tau = 329.27944\,$ms. Alternatively, without employing the correct value of $\alpha_n$, we can use Eq.~\eqref{eq:est_beta_n_p} and find $329.279\,\textrm{ms} < \tau \le 465.67\,\textrm{ms}$; secondly, employing Eq.~\eqref{eq:est1} we find $273.033\,\textrm{ms} < \tau < 329.581\,\textrm{ms}$. These two intervals only have a tiny overlap region which contains the true value of $\tau$.

Analyzing the $f$- and $p$-modes of different NS models, we make very similar findings in the accuracy of the estimates of frequency and damping time. Overall, we can confirm that this method works very well, and even if the features of the phase of the amplitude ratio are incomplete, a quite reliable value for the damping time can be extracted based on the presented inequalities.

\end{document}